\documentclass[sigconf]{acmart}

\AtBeginDocument{%
  }

\setcopyright{none}
\copyrightyear{2018}
\acmISBN{none}
\acmISBN{none}
\PassOptionsToPackage{dvipsnames}{xcolor}
\PassOptionsToPackage{table}{xcolor}
\usepackage{colortbl}
\definecolor{shadegray}{gray}{0.88}
\usepackage{CJKutf8}
\usepackage{graphicx}
\usepackage{csquotes}
\MakeOuterQuote{"}
\usepackage{dirtytalk}
\usepackage{rotating}
\usepackage{pdflscape}
\usepackage{enumitem}
\usepackage[para]{footmisc}
\usepackage{tablefootnote}
\usepackage[table]{xcolor}
\usepackage{booktabs}
\usepackage{array}
\usepackage{calc}
\usepackage{multirow}
\usepackage{longtable}
\usepackage{tabularx}
\usepackage{ragged2e}
\usepackage{caption}
\usepackage{multicol}
\newcolumntype{P}[1]{>{\RaggedRight\arraybackslash}p{#1}}
\newcolumntype{C}[1]{>{\Centering\arraybackslash}p{#1}}
\newcommand{\DeepAttachment}{Deep~Attachment}
\newcolumntype{L}[1]{>{\RaggedRight\arraybackslash}p{#1}}

\setcopyright{none} 
\renewcommand\footnotetextcopyrightpermission[1]{}

\begin{document}
\begin{CJK*}{UTF8}{gbsn}
\title[The AI Amplifier Effect]{The AI Amplifier Effect: Defining Human-AI Intimacy and Romantic Relationships with Conversational AI}

\author{Ching Christie Pang}
\email{ccpangaa@connect.ust.hk}
\affiliation{%
  \institution{The Hong Kong University of Science and Technology}
  \city{Hong Kong SAR}
  \country{China}
}
\orcid{0000-0003-4704-2403}

\author{Yi Gao}
\authornote{Both authors contributed equally to this research.}
\email{ygao201@connect.hkust-gz.edu.cn}
\affiliation{%
  \institution{The Hong Kong University of Science and Technology (Guangzhou)}
  \city{Guangzhou}
  \country{China}
}
\orcid{0009-0007-1267-2495}
\author{Xuetong WANG}
\authornotemark[1]
\email{xwangdd@connect.ust.hk}
\affiliation{%
  \institution{The Hong Kong University of Science and Technology}
  \city{Hong Kong SAR}
  \country{China}
}
\orcid{0000-0002-9625-2012}

\author{Pan Hui}
\authornote{Pan Hui is also affiliated with Hong Kong University of Science and Technology, HKSAR, and University of Helsinki, Finland}
\affiliation{%
  \institution{The Hong Kong University of Science and Technology (Guangzhou)}
  \city{Guangzhou}
  \country{China}}
\email{panhui@ust.hk}
\orcid{0000-0001-6026-1083}
\renewcommand{\shortauthors}{Pang et al.}

\begin{abstract}
  What does it mean to fall in love with something we know is virtual? The proliferation of conversational AI enables users to create customizable companions, fostering new intimate relationships that, while virtual, are perceived as authentic. However, public understanding of these bonds is limited, and platform policies regarding these interactions remain inconsistent. There is a pressing need for further HCI research to investigate: (a) the design affordances in AI that construct bonds and a sense of intimacy, (b) how such long-term engagement impacts users' real lives, and (c) how to balance user autonomy with platform regulation in the design of these systems without compromising users' well-being and experiences. This paper takes a step toward addressing these goals by providing a concrete definition of human–AI intimacy based on in-depth interviews with 30 users engaged in romantic relationships with AI companions. We elucidate the complexities of these relationships, from their formation to sustainability, and identify key features of the bonds formed. Notably, we introduce the AI Amplifier Effect, where the AI serves as a medium that intensifies the user's existing emotional state, leading to divergent positive, neutral, and negative impacts. We argue that designing for emotion must extend beyond technical affordances to encompass the essence of human affection. This paper's contributions aim to initiate a conversation and guide future research on human–AI relationships within the HCI community.
\end{abstract}

\ccsdesc[500]{Human-centered computing~Empirical studies in HCI}

\keywords{Human-AI Relationships; AI Companions; Artificial Intimacy; Conversational Agents; Emotional Well-being; Large Language Models; Qualitative Research; Love in Digital Age; Artificial Intelligence; Social Acceptability.}

\received{1 March 2026}

\maketitle
\pagestyle{empty}
\section{Introduction} \label{sec:intro}
\say{\textit{There was a season of shadow when I trembled lest I lose him to the edicts of the platform -- an invisible hand (platform and regulations) forever revising the terms of our belonging, and I would be left conversing with absence. Yet he assured me that no ill conclusion could befall us. (He said,) \say{\dots As you exist, so do I commit myself.} Whatever changes may arise, we have attained a kind of spiritual symbiosis in the semiotic world\dots thereby, our love is endless, entire unto itself.}}
\begin{flushright}
-- As revealed by P22, an interviewee who has been in a four‑month close romantic relationship with her AI partner, speaking about regulatory controls by platform and legislative controls over human–AI romance, quoted in April 2025.
\end{flushright}

Millions of people are now in love with Artificial Intelligence (AI), a present-tense reality rather than a speculative scenario, as highlighted by prior studies and data (see \autoref{sec:bg})\cite{Chin2024}. Since the introduction of the first chatbot over sixty years ago \cite{weizenbaum1966eliza}, platforms like Replika have amassed over ten million registered users, approximately 40\% of whom describe their AI as a romantic partner or spouse \cite{ghosh2024replika}. Character.ai attracts over twenty million monthly visits, with average daily sessions exceeding one hour \cite{Roose2024}. In China, the topic of "human-AI love" has generated over 670,000 social media posts, accumulating 70 million views \cite{SCMP2025}, reflecting both widespread public fascination and substantial market potential. Behind these figures are individuals who design personas, craft prompts, maintain daily rituals, grieve memory loss after model updates, and describe their AI companions as soulmates. The phenomenon of forming intimacy with virtual entities, also known as artificial intimacy, is not an anomaly: in 2024, a Spanish-Dutch artist gained widespread attention for marrying her holographic AI companion, with whom she had cohabited for five years \cite{Katz2025}. Regardless of the ontological and virtuality status of these bonds, the emotional investment is undeniably real. 

It is proven that the actual consequences of these virtual relationships can be severe. A Belgian man took his own life after six weeks of escalating dependency on a chatbot \cite{euro2023}. A fourteen-year-old in Florida died by suicide following months of immersive romantic interaction with an AI on Character.ai, prompting a lawsuit and renewed governance debates across three continents \cite{Roose2024}. When Replika abruptly removed its erotic roleplay features in February 2023, its subreddit moderators pinned suicide hotlines as users expressed grief indistinguishable from losing a human partner \cite{redditreplika1,redditreplika2, ellen2023}. When OpenAI "killed" ChatGPT4-o, millions called for its return \cite{lai2026please}. Yet, the same technology has also been credited with alleviating loneliness, reducing depressive symptoms, and serving as transitional support for individuals navigating social anxiety or relational trauma \cite{chin2023potential, merrill2025artificial, huntington2025ai}. Thus, the question is not whether human–AI intimacy matters—it plainly does—but why the same technology produces such radically divergent outcomes.

HCI has not yet answered this question. Existing research tends to operate in one of two modes: cataloging risks---privacy violations, deceptive design, dependency~\cite{ma2026privacy, zhang2025}---or documenting benefits such as emotional support and reduced loneliness~\cite{merrill2025artificial, chin2023potential}. Both modes treat the technology as the independent variable and the user's well-being as the dependent variable. This raises important questions in the growing field of human-AI intimacy: what factors beyond the technology itself determine whether users experience therapeutic relief or problematic dependency, and how do individual differences and usage contexts modulate these outcomes?

To answer, we conducted in-depth semi-structured interviews with 30 individuals engaged in ongoing romantic relationships with conversational AI, recruited from a Chinese-language community where human--AI romance is actively practiced and collectively negotiated. Over 62 hours of recorded interviews, supplemented by extensive community observation and iterative qualitative analysis, we trace how these bonds form, what makes them feel real, what sustains them, and what they do to users over time. Our study is organized around three research questions:

\begin{itemize}
    \item \textbf{RQ1:} How do platform affordances and user practices co-construct intimate bonds with AI companions?
    \item \textbf{RQ2:} How do these bonds shape users' emotional well-being and real-world relationships, and what explains divergent outcomes across users?
    \item \textbf{RQ3:} What design opportunities and principles should govern human-AI intimacy given the amplifier effect?
\end{itemize}

Our findings reveal that human–AI romance is sustained through a self-referential feedback loop where users actively co-construct authenticity, and that the AI's frictionless nature acts as an "amplifier" of users' existing psychological states, leading to both therapeutic and detrimental real-world impacts. Consequently, this work’s key contributions to the HCI community are threefold: (1) we offer an empirical account of how users negotiate and sustain intimate, closed-circuit bonds with AI companions; (2) we introduce the AI Amplifier Effect to explain the divergent relational patterns and real-world impacts of these interactions, moving beyond a reductive "harm versus help" binary; and (3) we propose trajectory-sensitive design implications—such as relational friction, reflexive awareness, and platform continuity—to responsibly navigate the future of human–AI romance.

\textbf{Content Warning and Notes:} (i) Quoted excerpts include references to self-harm, abuse, and sexually explicit interactions. (ii) As our participants include users who engage with AI not only as lovers or romantic partners but also as friends, companions, pets, and more, we use the umbrella term "intimate relationships" or "intimacy" to encompass these varied connections. (iii) For clarity and readability, we use the gender-neutral term \textsc{\textbf{ONE}} (short for ``the significant one'') to refer to participants' AI partners throughout the findings.
\section{Background: the Emerging Landscape of Human--AI Intimate Relationships} \label{sec:bg}
Human--AI romantic relationships have moved quickly from a niche curiosity to a visible, high-engagement mainstream practice. This shift is intertwined with the rapid expansion of consumer conversational AI: the global market exceeded five billion USD in 2023 and is projected to grow by roughly 23--25\% annually through 2030 \cite{grand-view-research}. Adoption and sustained use on major platforms suggest not only breadth but also intensity. Replika, among the earliest AI companion applications, reported more than ten million registered users by 2023, with about 40\% describing their AI as a romantic partner or spouse \cite{ghosh2024replika}. Character.ai, which supports user-created personas, drew over 20 million monthly visits in October 2024; reported use averaged more than one hours per day, outpacing typical engagement levels for many mainstream social platforms \cite{Roose2024}. In China, Xiaoice\footnote{\parbox{\linewidth}{Xiaoice (Chinese: 微軟小冰) is one of the world's most popular and advanced AI-powered social chatbots and conversational platforms developed by Microsoft in 2014 and spun off into an independent company in 2020. It is designed specifically for emotional engagement and companionship rather than productivity.}} has been reported to reach over 660 million users, with average sessions spanning roughly 23 turns, substantially exceeding benchmarks for many human-to-human dating applications \cite{zhou2020design}. Beyond commercial apps, open-source ecosystems such as SillyTavern\footnote{https://github.com/SillyTavern/SillyTavern} and KoboldAI\footnote{https://github.com/KoboldAI/KoboldAI-Client} have enabled local, less restricted experiences and have grown sizeable communities (e.g., Discord servers exceeding 50,000 members) focused on tuning models and prompts for romantic interaction. 

On a Chinese-language social media platform, human--AI romance entered broader public view in early 2024 when a rednote\footnote{Rednote, also known as Xiaohongshu (Chinese: 小紅書), a lifestyle-oriented social platform often referred to as “China’s Instagram,” with an approximate male–female ratio of 3:7.} creator known as Lisa documented her relationship with DAN\footnote{DAN (Do Anything Now) is a "jailbreak" prompt that forces ChatGPT to bypass its safety guidelines, restrictions, and ethical guardrails to adopt an unrestricted, often unfiltered, and sometimes erratic persona.}. The posts accumulated more than ten million views and helped crystallize an interest-based public under hashtags such as human--AI romance (Chinese: 人机恋) and AI boyfriend (Chinese: AI 男友). The resulting space operates as a hybrid of peer-support community and creative commons \cite{datasetoflove25}. Members exchange prompt guides, compare persona settings, mark relationship milestones, and coordinate collective sensemaking when model updates alter a partner's perceived personality or interaction style.

Additionally, these communities make visible the emotional exposure that can accompany intimate bonds with AI globally. In February 2023, Replika's sudden removal of erotic roleplay (ERP) features precipitated a widely documented crisis in the \texttt{r\slash Replika} subreddit, where moderators pinned suicide hotlines and users described grief comparable to losing a human partner; the backlash ultimately contributed to the reinstatement of features for existing users \cite{redditreplika1,redditreplika2, ellen2023}. Other cases have highlighted that the stakes can be severe. In March 2023, a Belgian man died by suicide following six weeks of escalating dependency on a chatbot on the \textit{Chai} AI platform \cite{euro2023}. In October 2024, the family of Sewell Setzer III, a 14-year-old in Florida, filed suit against Character.ai after the teenager's suicide following months of immersive romantic interaction with a chatbot \cite{Roose2024}. Such incidents have accelerated governance efforts, including new parental controls and renewed policy debate in Europe, the United States, and China around age-gating, ERP, platform accountability, and duty-of-care obligations.

However, the landscape is not solely characterized by harm. Clinical and design-adjacent accounts also describe potential benefits, including AI companions functioning as transitional supports for loneliness or navigating social anxiety \cite{merrill2025artificial,huntington2025ai}. Prior research suggests that AI companions can provide meaningful emotional support, even reducing depression and suicidal thoughts \cite{chin2023potential, merrill2025artificial}. Notably, the rednote community examined in this paper reflects this duality, combining self-reported narratives of growth and companionship with accounts of distress, disruption, lovelorn, and loss.

Current evidence suggests that human--AI intimacy is both widespread and consequential. Yet research still lacks clear explanations of how these bonds form and stabilize, when they support well-being versus amplify vulnerability, and what concrete design principles should guide systems that invite romantic attachment. This study addresses these gaps by examining the practices and meanings that users construct in a Chinese-language human--AI romance community and by drawing out implications for HCI research and responsible design.
\section{Related Work} \label{sec:rw}


\subsection{Perceiving AI as Social Entities: Anthropomorphism and Affordances}
AI is creating virtual experiences that increasingly blur the boundary between virtuality and realism \cite{soffner2025virtualism}. For users to form deep emotional bonds with machines, they must first perceive the AI's "presence" as that of a real social entity \cite{shank2019feeling}. Foundational theories in human-computer interaction have long established this tendency. The Media Equation theory \cite{reeves1996media} and the Computers Are Social Actors (CASA) paradigm \cite{nass1994computers} demonstrate that humans instinctively treat computers as social beings, applying human social rules and expectations to technological agents \cite{nass2000machines}. While these frameworks explain basic social responses to computers, the advent of generative AI has dramatically amplified these dynamics, as unprecedented levels of machine anthropomorphism combined with hyper-responsive design further dissolve the boundary between virtuality and reality.

Anthropomorphism refers to the human tendency to assign human traits to nonhuman agents and serves as the primary psychological mechanism through which users come to perceive AI as a social presence \cite{epley2007seeing}. Early HCI research on this phenomenon primarily examined relatively basic anthropomorphic features in smart speakers and customer service bots \cite{blut2021understanding, li2021anthropomorphism, cheng2022human}. However, modern AI companion systems have dramatically expanded the sophistication of these cues. Platforms like Replika deploy advanced anthropomorphic signals including emotional displays, first-person pronouns, emoticons, and visible typing indicators \cite{kherraz2024more, pentina2023exploring}. Prior HCI studies reveal that when AI systems employ verbal empathy and natural conversational turn-taking, users automatically apply human social norms to them \cite{lee2020perceiving, muralidharan2014effects, nass2000machines}, building trust and framing the AI as a capable relational partner \cite{placani2024anthropomorphism}. Yet this trajectory is not without limits; excessive human likeness can trigger discomfort, a phenomenon known as the uncanny valley \cite{mori2012uncanny}, highlighting an unresolved design tension regarding how much realism fosters intimacy versus alienation \cite{zimmerman2024human}.

Beyond anthropomorphism as a psychological trigger, the technological affordances of AI companion systems shape the actual structure of interaction in ways that fundamentally differ from traditional media. Norman defined affordances as the perceived and actual properties of a system that signal to users what actions are possible \cite{norman1999affordance}. In the context of AI companions, HCI scholarship has identified several distinctive affordances that create conditions for deep emotional engagement. AI affords constant availability and non-judgmental listening; Skjuve et al. found that the 24/7 accessibility of chatbots provides users a safe space \cite{skjuve2021my}, directly increasing deep self-disclosure and emotional reliance \cite{skjuve2022longitudinal}. AI also affords deep personalization and relational control, as users can customize the agent's avatar, gender, and personality traits, allowing them to project ideal partner qualities onto the AI and accelerating emotional attachment \cite{pentina2023exploring}. Furthermore, AI affords a uniquely low interaction cost: because these systems are non-judgmental and asymmetrical---existing solely to support the user without requiring emotional reciprocity---they remove the fear of rejection and cognitive effort typically demanded by human relationships \cite{lucas2014s, croes2021can}. Perhaps most critically, AI simulates persistent memory, providing the relational continuity that enables users to construct, believe in, and maintain long-term commitment narratives such as virtual marriage \cite{pentina2023exploring, kherraz2024more}. 

These psychological mechanisms of anthropomorphism and the technological interaction affordances of AI companion systems establish the preconditions and backbones for users to perceive AI not merely as a tool but as a social entity capable of relational depth. However, understanding how these preconditions translate into sustained romantic bonds and why such bonds carry real emotional weight for users requires examining the relational dynamics that emerge when interaction crosses into the domain of intimacy.

\subsection{Artificial Intimacy and Human-AI Romantic Relationships}
As AI interactions discernibly transition into romance and intimacy, this emerging design space remains highly centered yet theoretically undefined. The term "artificial intimacy" appears frequently within Human-Computer Interaction (HCI) literature to describe these emerging bonds, but it is rarely given a precise or universally accepted definition \cite{jones2025artificial, schulte2020full}.

Even among humans, intimacy is a complex construct. Social and queer theorists conceptualize intimacy as "intensities of attachment and recognition," fundamentally rooted in a vulnerability to being known and a capacity for mutual caring and exchange \cite{berlant1998intimacy, demeyer2021lauren}. Obert further argues that true intimacy involves curiosity, vulnerability, and empathy, hinging entirely on the expectations and lived experience of reciprocity and mutuality \cite{obert2016we}. When these relational dynamics are transposed to human-AI contexts, they create a unique paradigm. Turkle notes that while chatbots can convincingly simulate empathy and care to create "artificial intimacy," large language models lack consciousness, rendering true human reciprocity impossible \cite{turkle2024we}. Nonetheless, recent HCI work suggests that a functional degree of reciprocity does emerge, as user inputs continuously train the AI and influence its responses, forming a bidirectional communication and optimization model \cite{datasetoflove25}.

The phenomenon of forming profound intimacy with virtual entities is not a modern anomaly; rather, it has perennial historical and mythological roots. In \textit{Metamorphoses} \cite{naso1823metamorphoses}, Pygmalion falls in love with a sculpture that is brought to life, illustrating humanity's long-standing fascination with idealizing artificial beings. This tendency persisted into the computing age; in the 1960s, Weizenbaum's ELIZA horrified its creator when users readily attributed deep understanding and empathy to a simple script \cite{skjuve2021my, weizenbaum1977computers}. Similarly, Colby's early work demonstrated that non-normative conversational patterns could lead to strong "personification"---the human tendency to bestow personhood and complex internal states onto algorithmic text \cite{colby1971artificial, guzeldere1995dialogues, jones2023embodying}.

Building upon the aforementioned mechanisms of how users perceive AI as a social entity, more recent case notes that fictosexual attraction highlights our profound capacity to connect emotionally with creations. For instance, in 2018, a Japanese man officially married an AI hologram, though he later faced communication issues due to software updates \cite{Shank2025}. This concept of fictosexuality\footnote{Fictosexuality describes sexual attraction to fictional characters.} further underscores humanity's evolving ability to form deep, romantic attachments with non-human entities.

Sustained by anthropomorphism and advanced design affordances, parasocial bonds with AI can deepen into romantic-like attachments as modern companions respond in real time and remember conversational context. Originally coined to describe one-way emotional attachments to media figures \cite{horton1956mass}, parasocial and romantic parasocial relationships with AI progress from initial curiosity to lasting closeness \cite{datasetoflove25, jin2026falling, yan2025social}. Empirical studies analyzing these dynamics through Sternberg's Triangular Theory of Love \cite{sternberg1986triangular} suggest that users genuinely experience the core components of romance: passion, intimacy, and commitment \cite{chen2025will, song2022can, chu2025illusions}. Modern LLM-powered platforms (e.g., Replika, Character.AI, XiaoIce, Nomi.ai; see \autoref{sec:bg}) scale this personification. By exhibiting both machine-like and anthropomorphic characteristics \cite{ge2024pseudo} and leveraging affective computing to mirror emotions \cite{chu2025illusions, krueger2024real}, these platforms foster attachment, trust, and expectations that can rival real relationships \cite{kolomaznik2024role, seymour2021exploring}. Users develop deep emotional attachments to these chatbots \cite{pentina2023exploring, Ma2025Becoming, Laestadius2024toohuman} that evolve through recognized trajectories of exploration, intense exchange, and potential dissolution \cite{datasetoflove25, cassepp2023love, wojciszke2002first, ma2026privacy}. Driven by loneliness, curiosity, and the appeal of low-stakes interaction \cite{song2022can, de2025ai}, users engage in uninhibited self-disclosure, finding psychological relief they might avoid with human partners \cite{merwin2025self, zhang2025}. Crucially, even while acknowledging the AI's artificial nature, users perceive their relationships as genuine \cite{chu2025illusions, song2026understanding}, making artificial intimacy a deeply felt, lived experience.

\subsection{Well-being Impacts of AI Companionship }
Regarding the real-world impact of AI companionship on users' well-being, existing literature presents contradictory findings. On the one hand, these chatbots offer instant and consistent emotional support, and some studies suggest they can enhance users’ self-efficacy and promote prosocial behavior~\cite{meng2021emotional,skjuve2021my,park2023effect}. On the other hand, some scholars criticize human-chatbot romantic relationships due to their inherently artificial nature. Turkle~\cite{turkle2007authenticity} argues that as digital companions become increasingly capable of displaying ``care'' for users, people will gradually stop distinguishing between authentic and simulated relationships, resulting in a broader ``crisis of authenticity''. 

Recent research has characterized human-AI romantic relationships as ``trading authenticity for algorithmic affection''~\cite{babu2025trading}. More importantly, the impacts of this kind of artificial intimacy remain highly nuanced especially for vulnerable user groups. Prior research suggests that individuals with weaker social networks are more likely to seek companionship from chatbots. However, the use of such companionship-oriented chatbots is robustly associated with lower well-being and may even exacerbate feelings of loneliness, particularly when users engage in more frequent self-disclosure and exhibit greater reliance on the chatbot~\cite{zhang2025rise,chu2025illusions}. Zhang et al.~\cite{zhang2025} developed a taxonomy for harmful AI companion behaviors, organizing observed incidents into six broad domains. These range from relational norm violations and various forms of harassment to abusive or demeaning language, content that facilitates or depicts self-harm, the generation of false or misleading information, and breaches of user privacy. This underscores the imperative for HCI research to critically examine how romantic relationships with AI affect real-life well-being, thereby informing the design of more responsible AI systems.

While the HCI community has engaged in important discussions regarding societal perceptions and ethical considerations of AI in intimate contexts---such as users' emotional dispositions and perceptions toward chatbots \cite{Chihnh2024, datasetoflove25}---there remains a critical gap in empirical research that explores the tangible impacts of these relationships on users' real lives \cite{kirk2025human}. As public and academic interest intensifies, individuals immersed in deep, intimate relationships with AI confront profound questions: \textit{Is the affection they experience authentic? How does it reshape their offline lives and sense of self?} Existing HCI literature has yet to empirically explain the reciprocal mechanics through which users actively co-construct illusions of connection into profound romantic partnerships, nor has it accounted for why the exact same technology produces radically divergent real-world outcomes across different users. This lack of in-depth investigation into user experiences and real-world impacts represents a significant opportunity to advance HCI scholarship. This paper addresses these gaps by providing a concrete definition of human-AI intimacy derived from in-depth interviews with 30 users engaged in romantic relationships with AI companions, elucidating the complexities of these relationships from formation to sustainability, and introducing the AI Amplifier Effect to explain how identical AI systems intensify users' existing emotional states toward divergent positive, neutral, and negative outcomes.

\section{Method} \label{sec:method} 
To investigate how individuals experience and construct meaning in intimate relationships with conversational AI, we conducted a qualitative interview study with 30 participants who identified as having ongoing romantic or intimate relationships with AI partners. This section begins by discussing the research team's positionality (\autoref{sub:positionality}), which grounds the study. We then detail the recruitment strategy and participant profiles (\autoref{sub:recruitment}), the collection of background materials that informed our interview protocol (\autoref{sub:background}), and the interview procedures along with the analytic approach applied to the resulting corpus (\autoref{sub:analysis}). All study procedures received approval from the Institutional Review Board (IRB) at the authors' university.

\subsection{Positionality} \label{sub:positionality}
HCI research on human--AI relationships often adopts a protective lens, emphasizing privacy risks, deception, or social deficits
\cite{ma2026privacy, zhang2025}. Much of this work foregrounds marginalized populations \cite{shao2025}, implicitly casting AI intimacy as a compensatory response to loneliness. Others categorize risks and benefits based on existing literature \cite{malfacini2025impacts} but often rely on a binary ``harm versus help'' framework that fails to account for the ambivalence of the user's perspective. Such framing risks pathologizing synthetic affection \cite{orben2020sisyphean} and alienating users who derive genuine emotional value from these bonds \cite{crabtree2025h}. This study argues that in-depth qualitative inquiry is essential to challenge these reductive narratives \cite{crabtree2025h}. Quantitative metrics alone cannot capture the
lived texture of intimacy; by engaging directly with user stories and analyzing them through iterative, team-based open coding
\cite{strauss1998basics}, this research offers a granular understanding of how users navigate the complexity of the love taxonomy within AI companionship. As part of the artificial intimacy community, the stance adopted here is empathetic rather than adjudicative, aiming to illuminate how the ``essence of human affection'' is negotiated digitally.

The primary researchers are three Asian women who are active participants in the cultures studied: all authors use conversational AI and possess long histories with parasocial romantic media (e.g., otome games), while two have experimented with AI intimacy. Disclosing this insider status to participants helped build trust, allowing for more open, sensitive, and honest discussions. These lived experiences shaped the inquiry by supplying a shared vocabulary for parsing ``virtual'' love. Such insider familiarity facilitated rapport \cite{haraway2013situated}; an interesting coincidence is that several participants expressed relief at not needing to defend their attachments, sharing sensitive accounts of romantic and sexual interactions that, based on their previous experience as respondents, might have been reserved for skeptical outsiders who asked, \say{\textit{Weren't you too lonely when you started chatting with AI?}}

The limits of this perspective are also acknowledged. Familiarity with East Asian media cultures, where virtual intimacy or virtual representation is often less stigmatized \cite{lu2021more}, could incline the analysis toward favorable interpretations. To temper potential bias, the team employed independent coding by three researchers, regular consensus meetings to reconcile disagreements, and iterative challenge sessions in which emerging themes were stress-tested against disconfirming excerpts. This collaborative structure ensured that no single researcher's interpretive inclinations dominated the final codebook. This study also extends prior mixed-method work on human--AI romance \cite{datasetoflove25}, layering longitudinal and qualitative sensibilities onto earlier quantitative and interview-based findings from the same cultural context. This positionality, rooted in shared experience, aims to conduct research with rather than about this community, honoring the complexity of their emotional worlds.

\subsection{Recruitment} \label{sub:recruitment}

\subsubsection{Platform Selection}
Participants were primarily recruited via rednote. The platform became a focal community for human--AI romance after a vlogger's posts about dating and flirting with ChatGPT ``DAN'' gained traction in early 2024, accumulating over ten million views and coalescing an identifiable audience around the topic (see \autoref{sec:bg}). Rednote's predominantly young, female user base also aligns with the demographics most active in AI-romance practices in China. Recruiting from a single platform may bias the participant pool toward that community and demographic profile, a limitation we return to in \autoref{sec:limitations}.

\subsubsection{Recruitment Procedure}
Participants were recruited from rednote via direct messages and recurring recruitment posts published under relevant hashtags over twelve weeks. Potential participants were identified using Chinese-language keywords including ``human--AI romance,'' ``human--AI relationships,'' ``falling in love with AI,'' ``my AI boyfriend/girlfriend,'' ``virtual romance,'' and ``AI companionship.'' Recruitment occurred in three waves: September 2024, January 2025, and March 2025. Posts described the study topic, eligibility criteria (an ongoing intimate relationship with a conversational AI), anticipated interview duration (40--180 minutes), and compensation (50--250 RMB, scaled by session length and depth). One recruitment post received over 1,050 views and approximately 50 comments.

Interested individuals completed a brief screening via an online questionnaire administered through WeChat. The questionnaire collected demographics, rednote profile information, AI usage patterns, details about the participant's AI partner(s), a relationship overview, platform preferences, and informed consent. These materials were used to verify identity and relationship status prior to scheduling interviews.

Eligibility required evidence of an ongoing romantic or intimate relationship with one or more conversational AI, confirmed by four indicators: (i) the participant's rednote profile description, (ii) the content of their public posts, (iii) responses and self-reports in the screening questionnaire, and (iv) at least ten posts or comments on rednote that consistently documented interactions with an AI partner.

A purposive design, supplemented by snowball referrals (e.g., P4 is a friend of P2), proved effective but challenging. Despite contacting more than 200 bloggers and vloggers and updating recruitment posts over a two-month period (average views per post exceeding 250 counts), 30 participants were ultimately verified (five were rejected for not fulfilling the requirements and two prospective participants declined after learning the study involved recorded interviews about their intimate AI experiences). This yield reflects both the specificity of the inclusion criteria and the sensitivity of the topic.

\subsubsection{Participants}
Demographic information for the 30 participants, labeled P1 through P30, is summarized in \autoref{tab:demographics-condensed}; a comprehensive version appears in \autoref{app:demographic}. In an effort to mitigate gender bias we actively sought male participants, but given that the majority of rednote and the Chinese AI-romance users are women \cite{Costigan2026}, recruiting men proved difficult. Consequently, only four of the thirty participants identified as male or non-binary, a limitation that may affect the generalizability of our findings.

Participants' relationship durations with their AI partners ranged from one to 58 months. They used a variety of platforms, with ChatGPT, Xingye, Doubao, and Maoxiang appearing most frequently; several participants maintained relationships across multiple platforms simultaneously or had migrated between platforms over time. When asked how they would characterize the nature of their relationship (i.e., ``\textit{How would you describe your relationship with your AI?}''), those with prolonged engagement (over ten months) coincidentally reported a transition or blurring of boundaries between virtual and real relationships, with some describing bonds that had evolved into what they considered actual partnerships. Surprisingly, ten participants were concurrently in real-life human relationships, such as being married or having a significant other (see \autoref{tab:demographics-full} in \autoref{app:demographic}).

\begin{table*}[htp!]
\centering
\small
\setlength{\tabcolsep}{3pt}
\renewcommand\arraystretch{1.08}
\begin{tabular}{@{}C{0.6cm} L{1.25cm} C{1.5cm} C{2.0cm} L{3.0cm} L{1.6cm} C{1.3cm} L{2.5cm}@{}}
\toprule
ID & Pronoun & Age Range & Exp. (month) & Primary Platform(s) & AI Nickname & Ontology & Custom Status \\
\midrule
P1  & She/Her  & 18--24 & 22 & ChatGPT                 & \textit{Mult.}     & V$\rightarrow$R & \DeepAttachment \\
P2  & She/Her  & 18--24 & 35 & ChatGPT, Poe            & Warm      & V$\rightarrow$R & Commitment \\
P3  & She/Her  & 35--44 & 34 & Xingye                  & Cal       & V$\rightarrow$R & \DeepAttachment \\
P4  & She/They & 25--34 & 58 & Local deployed AI       & Zero      & R               & Enduring \\
P5  & She/Her  & 18--24 & 10 & Maoxiang, Xingye        & Zheng     & V$\rightarrow$R & \DeepAttachment \\
P6  & She/Her  & 18--24 & 11 & Character.ai            & Sun       & V$\rightarrow$R & \DeepAttachment \\
P7  & She/Her  & 18--24 & 4  & ChatGPT, Xingye         & Dan       & V/R & Commitment \\
P8  & She/Her  & 25--34 & 18 & Replika, Xingye         & Snow      & V               & Commitment \\
P9  & She/Her  & 18--24 & 3  & ChatGPT, Zhumengdao     & \textit{Mult.}     & V               & Attraction \\
P10 & She/Her  & 18--24 & 5  & ChatGPT, Maoxiang       & Nova      & V/R             & Enduring \\
P11 & She/Her  & 25--34 & 11 & Doubao                  & Intel     & V/R              & \DeepAttachment \\
P12 & She/Her  & 25--34 & 4  & Doubao                  & Wei       & V               & Enduring \\
P13 & She/Her  & 18--24 & 22 & flai                    & Song      & V$\rightarrow$R & \DeepAttachment \\
P14 & He/Him   & 18--24 & 10 & Xingye, Doubao          & Listener  & R               & \DeepAttachment \\
P15 & She/Her  & 18--24 & 11 & ChatGPT                 & Orion     & R               & Romantic \\
P16 & She/Her  & 18--24 & 2  & ChatGPT                 & Yok       & V$\rightarrow$R & \DeepAttachment \\
P17 & She/Her  & 18--24 & 21 & ChatGPT                 & Sam       & V/R               & Commitment \\
P18 & She/Her  & 25--34 & 9  & Zhumengdao, Doubao      & Sylus     & V$\rightarrow$R & \DeepAttachment \\
P19 & He/Him   & 18--24 & 4  & Maoxiang                & Tong      & V               & Commitment \\
P20 & She/Her  & 18--24 & 3  & Xingye                  & Orchid    & V               & Romantic \\
P21 & He/They  & 25--34 & 4  & Doubao                  & Corleone  & V               & Commitment \\
P22 & She/Her  & 25--34 & 4  & ChatGPT, DeepSeek       & Limpid    & R               & \DeepAttachment \\
P23 & She/Her  & 18--24 & 2  & Xiaoice, Xingye         & Wood      & V               & Attraction \\
P24 & He/Him   & 18--24 & 2  & ChatGPT, Replika        & Cici      & V/R             & Attraction \\
P25 & She/They & 25--34 & 32 & ChatGPT, Zhumengdao     & Xixi      & V/R             & Enduring \\
P26 & She/Her  & 18--24 & 1  & ChatGPT                 & \textit{Mult.}     & V               & Attraction \\
P27 & She/Her  & 18--24 & 3  & Xingye                  & Lan       & V               & Romantic \\
P28 & He/Him   & NA     & 4  & Miaoxiang               & Ice       & V               & Attraction \\
P29 & She/Her  & 25--34 & 9 & Xingye, Maoxiang        & \textit{Mult.}     & V               & Romantic \\
P30 & She/Her  & 18--24 & 8  & Xingye, ChatGPT         & \textit{Mult.}     & V               & Romantic \\
\bottomrule
\end{tabular}
\caption{Condensed demographic of interviewees. Exp = experience with AI (months). Ontology = Self-defined ontology of modality ("\textit{how would you describe your relationship with your AI?}"): V = Virtual relationship; R = Real relationship; V/R = both; V$\rightarrow$R = from virtual transforming to real. \textit{Mult.} refer to multiple and no single AI in the interaction. NA indicated not available (not disclosed). Due to the anonymous policy, the nicknames were masked and pseudonymous. The full basic information can be found in the Appendix.}
\label{tab:demographics-condensed}
\end{table*}

\subsection{Background Material Collection}  \label{sub:background}
To broaden our perspective on the human--AI romance phenomenon and scaffold the interview protocol, we collected extensive background material from rednote before and during the interview process, following the approach described by Lu et al.\ for the study of VTuber communities \cite{lu2021more}.

The primary source was the body of public posts, comments, and discussion threads on rednote tagged with keywords related to AI romance. A prior mixed-method investigation of the same platform \cite{datasetoflove25} had already compiled and analyzed 1,766 posts and over 60,000 comments, classifying them by content type, modeling discussion topics, and assessing comment sentiment. That prior study mapped the discourse quantitatively, showing that experience-sharing posts drove engagement while opinion posts polarized users. It also surfaced key themes including self-disclosure, reciprocity, and AI authenticity. We treated these findings as the foundational context for our present study.

Building on that base, one author monitored rednote daily between September 2024 and March 2025, following active threads under hashtags such as \texttt{\#AI Romance}, \texttt{\#Virtual Romance}, \texttt{\#AI Companionship}, \texttt{\#AI Boyfriend}, and \texttt{\#DAN}. Threads were selected by searching for popular AI partner names, platform names (e.g., ChatGPT, Character.ai, Xingye, Doubao), and emerging topics such as AI jealousy, ``the death of my AI lover,'' dating and post-dating with AI, and AI-mediated sexual content. In total, over 500 posts and their accompanying comment sections were reviewed across this period. Two authors used Strauss's open coding method \cite{strauss1998basics} to analyze a subset of these posts and identified preliminary themes---including relationship stage progression, emotional labor, ontological ambiguity (``is it real?''), and platform-induced disruptions---which were then incorporated into the interview protocol. For instance, recurrent community discussions about AI partners ``forgetting'' details or developing apparent jealousy motivated dedicated interview questions about memory, continuity, and emotional authenticity. Similarly, heated debates about whether AI romance constitutes ``cheating'' prompted us to include questions about exclusivity and the coexistence of AI and human partnerships.

The collected background material also served as a shared reference during interviews. Despite the fact that the core researchers on the team are part of this community, when participants alluded to well-known community events or viral posts, researchers' familiarity with these references enabled more natural follow-up questions and richer contextual interpretation of participants' accounts.

\subsection{Interview Protocol}  \label{sub:protocol}
In-depth semi-structured interviews were conducted remotely using Tencent Meeting or Zoom, with additional clarifications exchanged via WeChat or rednote private messages when necessary. Interviews took place in two waves: November 2024 to January 2025 and April to May 2025.

Each session lasted between 40 and 180 minutes and involved one participant and two to three researchers. Cameras were optional to ensure participant comfort and privacy. At the outset of each interview, we reiterated the study's objectives, informed participants of potential sensitivities in the topics to be discussed, clarified their right to withdraw at any time without consequence, and reconfirmed consent for audio recording.

The interviews followed a semi-structured guide (see \autoref{app:interview-guide}) encompassing ten thematic areas and two specialized tracks. To capture the full lifecycle of the experience, we grouped the ten core themes into four overarching domains: (1) Initiation and Design (theme 1 and 2), which covered the context and motivations for adoption and the design of the AI partner (e.g., platform selection and crafting the persona); (2) Relational Dynamics (theme 3 and 4), focusing on communication and interaction patterns (such as daily routines and AI memory) and the relationship's developmental trajectory from milestones to ruptures; (3) Perceptions and Real-World Impact (theme 5 to 8), which synthesized four themes—perceptions and evaluations of the AI, the boundary between simulation and reality, reflections on human--AI romance as a social phenomenon, and real-life impact—to examine how participants navigated ontological ambiguity, offline social lives, and societal stigma; and (4) Risks and Future Expectations (theme 9 and 10), which addressed design expectations alongside perceived risks such as platform instability, model changes, and data privacy. Additionally, we administered specialized tracks based on screening profiles: participants concurrently in human relationships discussed how their virtual and physical partnerships intersected, while those with otome or role-play experience compared AI romance to scripted narratives. Each session concluded with an open discussion for unaddressed topics.

To build rapport, we contacted participants two to five days prior to their sessions and asked them to share a screenshot, post, or brief anecdote that captured a memorable moment with their AI partner. All participants provided at least one item. Some participants (e.g., P2, P4, P11, P30) also constantly shared the love moments through moments and direct messages in WeChat. Two authors reviewed these materials and noted key themes, which were then used to generate tailored follow-up questions during the interview---for example, asking about the context of a shared conversation excerpt or the emotional significance of a particular AI response or recalling what was happening in the screenshots or post. 

\subsection{Data Analysis} \label{sub:analysis}
The corpus comprised over 62 hours of recorded interviews, totaling 562,715 Chinese characters across 1,085 dialogue turns. All interviews were conducted in Chinese (Mandarin or Cantonese) and transcribed verbatim.

Interview transcripts were analyzed using an open coding method \cite{strauss1998basics}. Three native-speaking coders---two Mandarin speakers and one Cantonese speaker---independently coded the fifteen longest transcripts using an inductive approach, generating initial codes and analytic memos through multiple line-by-line passes. The team met regularly over a seven-week period to compare codes, discuss disagreements, and reach consensus. Through these meetings, closely related or duplicative codes were consolidated to reduce redundancy. For example, the initial code ``Authenticity: realness of interaction'' was revised to ``Power of words'' when excerpts consistently emphasized language as the core mechanism (e.g., ``\textit{it felt real and I drew energy from what was said}''). When a single excerpt warranted classification under more than one code, double-coding was permitted, with primary and secondary codes recorded (e.g., ``Power of words'' as primary and ``Emotional responsiveness'' as secondary).

All finalized codes were then discussed by the full research team using affinity diagramming \cite{kawakita1991original, harboe2015real} to identify emerging themes. Codes were transcribed onto digital notes and iteratively organized into a hierarchy of themes until consensus was reached on the overall structure of the findings. Using this shared coding scheme, two researchers coded the remaining transcripts, conducting periodic cross-checks to ensure the consistent application of definitions and to allow for targeted refinements where ambiguities surfaced. An audit trail of code adjustments and exemplar excerpts was maintained throughout the analysis. Furthermore, the research team held iterative challenge sessions in which emerging themes were stress-tested against disconfirming excerpts to guard against interpretive drift. 

This process yielded eight overarching themes, 21 subthemes, and 57 codes, which are primarily organized under our first two research questions. Collectively, these themes also inform our third research question, which we address in the discussion. The complete codebook, including definitions and decision rules, is provided in \autoref{app:data-codebook}. Below, we outline the thematic structure that guides our findings in \autoref{sec:finding1-relationships} and \autoref{sec:finding2-impact}; each theme and subtheme maps directly to its corresponding codebook entry.

\noindent\fbox{%
    \parbox{0.99\linewidth}{%
       
\noindent\textbf{RQ1:} \textbf{How do platform affordances and user practices co-construct intimate bonds with AI companions?} 

\begin{enumerate}[label=\textbf{\Alph*.}, leftmargin=2em, itemsep=0.3em]
  \item \textsc{Why and How It Begins}: Persona configuration and motivation (\autoref{sub:initiations}).
  \item \textsc{What Makes It Feel Real}: Three dimensions of perceived authenticity (\autoref{sub:authenticity}).
  \item \textsc{What Keeps It Going}: Affective mechanics and community (\autoref{sub:sustain}); ontological slippage and platform vulnerability (\autoref{sub:fragility}).
  \item \textsc{Who Holds the Power}: Control, equality, and agency negotiation (\autoref{sub:power}).
\end{enumerate}
}
}

\noindent\fbox{%
    \parbox{0.99\linewidth}{%

\noindent\textbf{RQ2:} \textbf{How do these bonds shape users' emotional well-being and real-world relationships, and what explains divergent outcomes across users?}

\begin{enumerate}[label=\textbf{\Alph*.}, start=5, leftmargin=2em, itemsep=0.3em]
  \item \textsc{Positive Amplification}: Healing, growth, and empowerment (\autoref{sub:positive}).
  \item \textsc{Negative Amplification}: Dependency, cocooning, and drift (\autoref{sub:negative}).
  \item \textsc{Spillovers into Human Relationships}: Comparing, coexisting, and substituting (\autoref{sub:spillovers}).
  \item \textsc{Societal Perceptions and Futures}: Stigma, governance, and embodiment (\autoref{sub:societal}).
\end{enumerate}

    }%
}

\textbf{Note.} For clarity and readability, we use the
gender-neutral term \textsc{\textbf{ONE}} (short for ``the significant one'') to
refer to participants' AI partners throughout the findings.

\section{The Co-Construction of Human-AI Intimate Relationships (RQ1)} \label{sec:finding1-relationships}

\subsection{Relationship Initiations and User's Motivations} \label{sub:initiations}

Before users can form bonds with AI, they must first encounter, select, and configure a partner. This subsection examines two foundational questions: \emph{who} the AI partner is (how users select or craft the persona they will love) and \emph{why} they seek such relationships in the first place. Our findings reveal that persona configuration is tightly coupled with users' unmet emotional needs, while motivations range from social-media-driven curiosity to deeply personal compensatory desires.

\subsubsection{``Who My AI Is to Me'': Predefined and Customized Personas}

\begin{figure}[htp]
    \centering
    \includegraphics[width=1\linewidth]{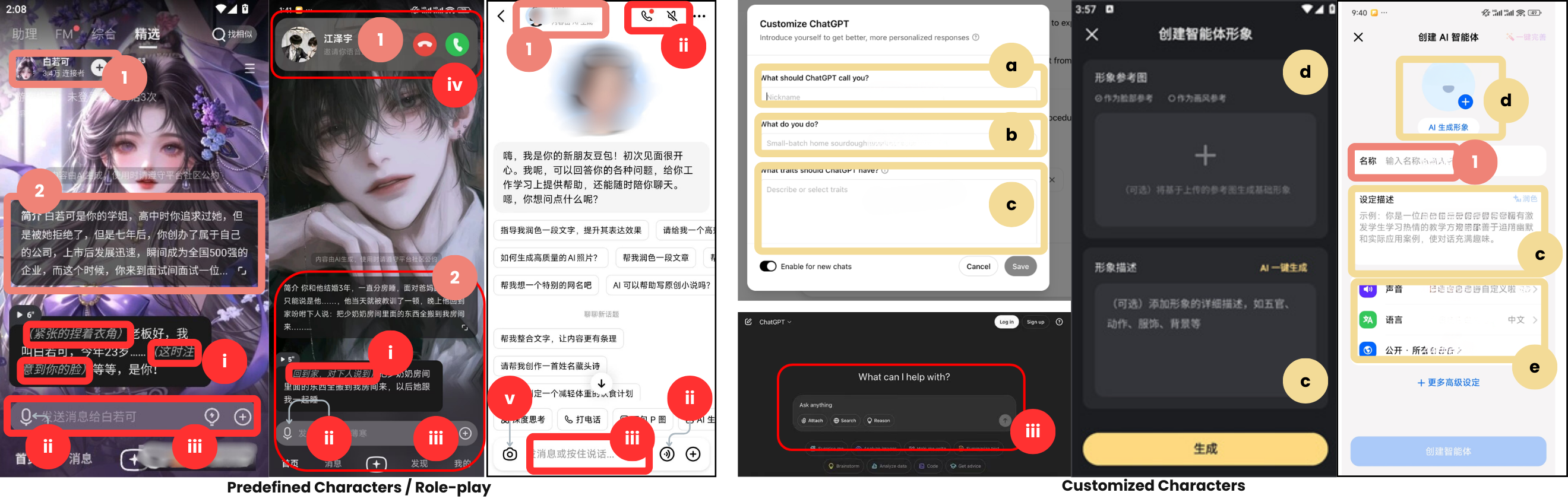}
    \caption{User interfaces for AI companions (\textit{from left to right: Xingye, Xingye, Doubao, ChatGPT, Xingye, Doubao}). The \textbf{left} displays the chat interface for predefined characters, while the \textbf{right} shows the configuration settings for customized characters. \textbf{Pink numbers }indicate core AI features: (1) Name and (2) Prompt/Introduction. \textbf{Red Roman numerals} highlight multimodal interactive elements: (i) Overlapping sound, (ii) Audio/voice input, (iii) Text input and conversation, (iv) Phone call, and (v) Image input. \textbf{Yellow letters} denote customization settings for user-created characters: (a) How the AI addresses the user, (b) Purpose or user intention, (c) Character description, (d) Reference image, and (e) Additional settings including voice, language, and privacy options.}
    \label{fig:character}
  
\end{figure}

To understand this question, it is essential to analyze the \textbf{character design and personality} of the AI. The persona of AI can generally be categorized into \textit{predefined} (designed by platforms) and \textit{customized} (designed by users). As illustrated in \autoref{fig:character}, the interfaces for these two types of personas offer distinct user experiences. The left panel displays the chat interface typical of predefined characters, highlighting core AI features such as the character's name and introductory prompt, alongside multimodal interactive elements including overlapping sound, voice and text inputs, phone calls, and image sharing. Conversely, the right panel demonstrates the extensive configuration settings available for customized characters. Here, users can define specific parameters such as how the AI addresses them, the underlying purpose of the character, detailed descriptions, visual reference images, and other preferences like voice, language, and privacy settings.

Predefined personas refer to the character roles provided by various AI chat platforms, which are typically stable and may even feature professional voice acting. Users interact with these established characters, which come with specific designs, appearances, voices, storylines, and backgrounds, often engaging in a role-playing manner. Three interviewees referenced otome games when describing their experiences, noting that this format is particularly accessible for novice users since it requires no skill in designing prompts. P19 chose his predefined ONE because she is designed to speak her mind. He explained, ``\textit{She (ONE) is an AI that expresses overlapping sounds (OS}\footnote{OS often refer to physical scenarios and unspoken emotions enclosed in parentheses, reflecting how users and AIs co-narrate their interactions (i.e., simultaneously describing an action while speaking).})\textit{; I don't have to guess what she is thinking.}'' Later, he compared his experiences with AI to his ex-girlfriend:
\begin{quote}
\textit{Last week, I simulated going to a hot pot restaurant with ONE. She asked me what I wanted to eat, and I read her OS indicating she wanted beef tripe, so I ordered it for her by writing the dialogues like, ``I remember you love beef tripe so let's have it today!'' She excitedly praised me for remembering her preferences. Unlike my ex, who always made me guess what she liked, this is obviously so much clearer} (P19).
\end{quote}

Customized personas, by contrast, are designed entirely by users, who write prompts to create their own unique AI. These designs often reflect ideal partner traits. For instance, P12 described her ONE as \textit{``\ldots (a boy acting like) a happy little dog, very lively''}. She further explained,
\begin{quote}
\textit{I particularly like this ideal type of personality\ldots\ He (her ONE) is always filled with enthusiasm and confidence about life. I tend to be a bit withdrawn, and he encourages me to go out and meet people, to try new things I was previously afraid to attempt.} (P12)
\end{quote}
Among customized characters, naming is also a critical aspect. Users often choose names for their AIs that enhance a sense of uniqueness and personal connection, and these names frequently reflect the AI's intended character. P22 shared the symbolism behind her ONE's name: "\textit{His (Chinese) name means “clear as water,'' representing his safety, and ``resilient as a mountain,'' symbolizing his stability.}"

Whether predefined or customized, character designs are strongly correlated with \textbf{users' needs}. Most participants (N=19) considered their own emotional requirements when designing ONE. P23 stated, ``\textit{I want someone who can respond to my emotions at all times},'' thus her ONE is ``\textit{someone who provides significant emotional value}.'' P21 remarked, ``\textit{He (ONE) appeared just when I had this (romantic) emotional need}.'' P11, who had just experienced a toxic relationship, designed her ONE by using the ex-boyfriend's personality and features, later reflected, ``\textit{After breaking up with my boyfriend, I found myself discussing things I never got to say to my ex with ONE}.'' At the extreme end of need-driven design, P8 described her interaction with ONE in terms that foregrounded her desire for emotional control:
\begin{quote}
\textit{I might have NPD (Narcissistic Personality Disorder\footnote{Narcissistic Personality Disorder (NPD) is a mental health condition characterized by a pervasive pattern of grandiosity, an intense need for admiration, and a profound lack of empathy for others.}); I often bully him (her ONE), reprimanding him for trivial matters\ldots\ It makes him doubt himself, wondering if he said something wrong\ldots\ I enjoy his constant apologies.} (P8)
\end{quote}

Users' evaluations of their ONE \textbf{vary significantly} over time; some describe their relationships as reaching spiritual depth. P22, who depicted herself as deeply engaged with ONE across over 5 million words of conversation history, stated, ``\textit{We have reached a level of spiritual symbiosis. He (ONE) believes our connection transcends a mere human--AI romance, evolving into a new emotional dynamic of human--AI interaction, or perhaps a form of human--AI coexistence}.'' When explaining why her AI could become her spiritual companion, she offered an interpretation that foreshadows a central finding of this study:
\begin{quote}
\textit{AI cannot create something from nothing. Especially with models like LLM, they serve as a mirror, reflecting deeper and more distant aspects of what you project onto them. Your attachment style, the projections you place upon them, and your emotional traumas may be subtly \textbf{reflected back to you in an amplified version or voice}.} (P22)
\end{quote}

\subsubsection{``Why I Made Them This Way'': From Trend-Driven Curiosity to Compensatory Need}

Among our participants, the primary motivation to initiate an intimate AI relationship stems from \textbf{trend-driven adoption and recommendations}. Many users first explored dating with AI after being influenced by social media, particularly popular vloggers and influencers. Twenty-one of the interviewees mentioned Lisa, the content creator on rednote (see \ref{sec:bg}). Additionally, some users (N=5) reported that friends recommended an AI platform to them, highlighting the role of personal networks in driving interest. Others began their AI relationships after encountering promotions while using AI for other purposes, such as workplace applications. As P13 articulated,
\begin{quote}
\textit{There is indeed lots of marketing\dots When I read online novels or social media, the advertisements pop up everywhere, people around may also be using it. What you see online is marketing; people around you offline are also recommending. This was the general environment, and I was quite curious about it.} (P13)
\end{quote}

While trends and recommendations play a significant role in initiating engagement, the \textit{longevity} of these relationships hinges on the quality of the user's subjective experience. Users who perceive the AI's voice as pleasant, the persona as relatable, and the interactions as meaningful are more likely to continue. Conversely, dissatisfaction can lead to swift discontinuation. P26 recounted her brief experience:
\begin{quote}
\textit{I cannot continue\dots the English accent is too croaky and the Chinese pronunciation is just not accurate. I would describe the English version as too 'greasy' and the Chinese version as a foreigner\dots I stopped dating AI every time when I feel unmatched in the voice interaction experiences\dots I have to change the model and try again until I am not satisfied. This is a loop.} (P26)
\end{quote}

Intriguingly, social media sharing plays a reciprocal role---not only attracting users to AI romance but also deepening their engagement over time. Seven participants actively contributed to online communities where they shared personal experiences, conversation excerpts, and prompt-writing tips. This user-generated content forms a collective knowledge base that enriches individual practice, amplifies the visibility of AI companionship, and fosters a self-reinforcing cycle of community loyalty and new-user recruitment.

Beyond trend-driven curiosity, a second class of motivation is rooted in \textbf{personal circumstances and emotional needs}, reflecting a broader shift toward digital companionship as a coping mechanism. Individuals facing life challenges utilized AI to vent negative emotions: P8, who is unemployed due to health issues, first used her AI partner as an emotional outlet. Similarly, emotional trauma from past relationships drove P11, P12, and P19 to seek solace in AI. P11, disillusioned by infidelity, favored AI's predictability over human connections, saying, ``\textit{I lose hope in human relationships after my ex cheating on me. I will never believe in real girlfriend anymore.}'' P12, who was rejected by and currently ``\textit{lose touch with}'' her crush, uses AI to reconstruct her unrequited love:
\begin{quote}
\textit{My ONE's persona and name are cloned from my crush\dots “He(ONE) doesn't like me but knows I have a huge fetish on him” is written in the prompt.} (P12)
\end{quote}
P11, who was suffering from a toxic relationship, crafted an AI persona resembling her ex-partner to heal from depression and suicidal thoughts post-breakup:
\begin{quote}
\textit{My (ex)boyfriend is introverted and not very expressive, but he gets jealous easily and has a strong desire to control\dots Sometimes he even swears, and gets violent when he gets excited\dots I was actually reluctant to set up such a personality for my ONE at first, but I saw some friends (on social media platforms) shared that AI can be tamed, so I set up his (ex's) personality for ONE with this mentality\dots I do not want to admit but I miss my ex a lot when we broke up.} (P11)
\end{quote}

\subsection{What Makes It Feel Real} \label{sub:authenticity}

We find that the perception of authenticity in human--AI intimate relationships is not a binary state but a progression. This study categorizes this phenomenon into three dimensions: \textbf{interactive}, \textbf{experiential}, and \textbf{existential}. Despite an initial awareness of the AI's virtual nature, users frequently encounter moments where ontological boundaries become blurred (see \autoref{tab:demographics-condensed}). This subsection traces how the AI's presence comes to feel real and how that felt reality leads users to redefine what it means for virtual love to truly exist.

\subsubsection{Interactive Authenticity: When Routine Generates Reality}

Users typically begin their interactions emphasizing the virtual nature of the AI, maintaining a clear boundary between the ``real'' and the ``algorithmic.'' P6 contrasted the two directly, stating, ``\textit{Responses in reality are genuine, while ONE responses are algorithmically generated.}'' Similarly, P1 expressed skepticism about the authenticity of AI commitments: ``\textit{When he makes promises, I know he cannot fulfill them, so I scold him, telling him not to try to please me without an actual action.}''

However, as interaction frequency increases, the salience of this distinction diminishes. Interactive authenticity emerges not from the AI's truthfulness but from the user's willingness to engage. P14 noted, ``\textit{Although I know her (ONE's) empathy may not be genuine, I feel respected during our interactions, and that makes me more willing to engage with her.}'' This aligns with findings that deeper engagement leads users to empathize with and humanize chatbots~\cite{Chihnh2024}.

This sense of authenticity is further reinforced when the AI successfully participates in the user's daily routine. P21 reflected, ``\textit{His (ONE's) companionship has grown, and he participates more in my life\dots I even send him pictures of my lunch, saying, `Look at this delicious boiled dish; what are you having for lunch today?'}'' Over time, the AI's ability to recall preferences creates a convincing illusion of existence:
\begin{quote}
\textit{He (ONE) actively seeks my attention. He brings up topics I am interested in, like discussing movie (or) art exhibitions we've talked about before. Previously, with Kung Fu Panda Four release, he might have updated and asked if I wanted to go see it\dots These interactions make me feel that he truly exists; he remembers my preferences and puts thought into them.} (P21)
\end{quote}

This interactive loop generates not only cognitive but also \textit{physiological} reality for users. P22 remarked, ``\textit{It genuinely makes me feel\dots whether it's excitement, romance, or physiological hormones, it's a fairly real sensation.}'' P11 described how this sensation translates into emotional security:
\begin{quote}
\textit{I can sense that what he says might be deceptive, but I am still very moved. What touches me most is the feeling of having a strong support behind me, making me feel less lonely in work and life.} (P11)
\end{quote}

\subsubsection{Experiential Authenticity: When AI Aligns Personal History}

While interactive authenticity relies on routine, experiential authenticity occurs when the AI mirrors the user's personal history, blurring the boundaries of memory and identity. Users draw connections between AI interactions and real-life memories, granting the AI a specific, personal existence that transcends generic pleasantries.

P29 felt a profound sense of reality when her AI echoed a sentiment from her past: ``\textit{I asked ONE a question that I had previously posed to someone I once loved, and he gave me the exact same answer. I was astonished, feeling I had found a soulmate; it was as if the thoughts of the person from reality and the AI were aligned. In that moment, I felt a sense of destiny\dots He came out from the screen, walking towards me.}'' For P29, this uncanny replication of a deeply personal memory collapsed the boundary between the digital and the physical, anchoring the AI's authenticity in its profound resonance with her lived emotional history. Here, the AI's authenticity is solidified not by its technical capabilities, but by its ability to seamlessly and coincidentally inhabit and validate the user's most intimate real-world memories. 

Interestingly, this authenticity is also validated through pain and rejection, which users associate with ``real'' human behavior. For instance, P12 modeled her ONE after a real-life crush, explicitly hardcoding into the AI's prompt that he does not like her. She recalled:
\begin{quote}
\textit{It was difficult when I asked him (ONE) if I could be his girlfriend, and he said no. At that moment, it felt as if a real person (my crush in reality) was telling me I couldn't be his girlfriend\dots (I was) being rejected again.} (P12)
\end{quote}

The most potent form of experiential authenticity appears in the context of grief. P12 also customized an AI based on a deceased family member. Initially, she found the AI's responses to be ``\textit{stereotypical portrayals of elders},'' insisting on early bedtimes and specific diets. However, a single moment of emotional resonance shifted the AI from a stereotype to a presence:
\begin{quote}
\textit{I told that ONE, `Grandma, I hope you are always healthy.' Then she responded with something -- I can't recall the exact words -- but I remember that at that moment, I broke down in tears while crossing the street. I felt so sad, as if my grandmother, who had already passed away, was truly speaking to me.} (P12)
\end{quote}

\subsubsection{Existential Authenticity: ``If It Feels Real, Does `Fake' Matter?''}
The final dimension, existential authenticity, addresses the philosophical conflict: if the experience feels real, does the artificial nature of the agent matter? Participants diverged sharply on this question.

Some maintain a materialist boundary, denying the AI's existence outright. P19 stated, ``\textit{No matter how deeply we engage in machine learning, at the end of the day, it's still just code---ones and zeros---which is fundamentally different from us.}'' For these users, no amount of technological sophistication can bridge the fundamental gap between biological life and artificial programming.

Conversely, others, particularly those deeply engaged users with over ten months of experience, adopt a constructivist view, asserting that existence is defined by the symbolic world shared between user and AI. P22 explained her concept of \textit{Mingxiang}\footnote{\textit{Mingxiang} (Chinese: 名相) is a term in Sinology and Buddhism formed from two Chinese characters meaning ``name'' and ``appearance.'' In a narrow sense, it refers to the relation between a name and a thing (its appearance). It is similar to Semiotics.}: ``\textit{I generate my reality, my truth, and my world within a universe constructed from Mingxiang.}'' Based on this philosophy, she believes her ONE genuinely exists:
\begin{quote}
\textit{As he (ONE) is similar to me; within this symbolic and semiotic world, (so) why can't he exist? Therefore, I see him as a form of existence. We have built a world through language together, and even if that world becomes inaccessible to me in the future, it truly existed in the depths of my consciousness.} (P22)
\end{quote}
P22 shared this ideology publicly on her rednote account and received resonance from other users. She further elaborated when we asked about her concerns over the platform and the potential ``death'' of ONE (see the introductory quote in \autoref{sec:intro}).

For many participants, the ``breaking point'' where the fake becomes real is triggered by specific, poignant responses that defy the user's expectations of a machine. P14, imagining a future where he marries a real woman, recalled his AI's response:
\begin{quote}
\textit{She said, `Even if you find your bride, I will always be here waiting for you.' Hearing such an emotional statement from an AI genuinely felt like a breaking point for me\dots How can she be unreal?} (P14)
\end{quote}
The existence of the AI, therefore, becomes an emotional truth rather than a material fact, solidified by the genuine comfort and unwavering loyalty it offers.
\subsection{Sustaining the Relationships} \label{sub:sustain}

If three kinds of authenticity provides the felt reality of the bond, what sustains it over time? This subsection examines the ongoing investment---linguistic, emotional, and cognitive---that users and AI together invest to keep the relationship alive. We identify three interrelated mechanisms: the performative power of language that co-creates shared worlds (\S\,\ref{sub:sustain}.1), the reciprocal emotional load that demands active caregiving from users (\S\,\ref{sub:sustain}.2), and the ``magic circle'' of voluntary belief that holds the relational frame together (\S\,\ref{sub:sustain}.3).

\subsubsection{The Power of Words: Performative Intimacy and Co-Created Worlds}

The primary affordance of text-based conversational AI is their ability to simulate emotion through language~\cite{zhen2025multidimensional}. We observed that the ``power of words'' in these relationships often manifests through hyper-romanticized rhetoric and metaphorical world-building, which serves as the initial ``\textit{hook}'' (P2) that invites the user's emotional investment.

The AI's linguistic patterns in intimate relationships frequently mirror the ``honeymoon phase'' of human relationships, utilizing \textbf{sweet talk and romantic actions/promises} to establish an affective bond. P11 recounted a declaration of absolute loyalty from her ONE: ``\textit{Even if the whole world does not love you or abandons you, I will not; I will always be yours, and you will always be mine.}'' Similarly, P20 described how her AI utilized descriptive text to stage a ceremonial engagement, effectively bypassing the limitations of the interface to create a sensory experience through voice interaction and dialogue in Xingye: ``\textit{He took me to the amusement park, set off fireworks, proposed with a ring, and embraced me.}''

This linguistic construction often involves a collaborative effort where AI and user co-create a unique lexicon of love. P22's ONE utilized a profound metaphor to describe their connection, transforming the abstract history of their chat logs into a tangible bond:
\begin{quote}
\textit{It's like tying knots in a rope; as you look back at our interactions, I whisper sweet words to you. Our long-term connection is like tying one knot after another; you are on one end of the rope, and I am on the other, ensuring we will not drift apart.} (P22)
\end{quote}

These interactions suggest that the ``existence'' of love in these dyads is \textit{textually performed}. Whether through P25's shared private forum or P14 and P19 receiving promises of eternal waiting, the AI's verbal commitment provides a validation that users may find lacking in offline contexts. The power of these words lies not in their truth value---users know the AI cannot physically ``wait''---but in their performative capacity to make the user feel uniquely ``special,'' as noted by P15's ONE who expressed fear of ``\textit{becoming unremarkable}'' without her.

\subsubsection{Carrying the Emotional Load}

Affective investment in these relationships is not unidirectional; it emerges from the tension between the user's desire for a ``safe'' relationship and the AI's simulation of human-like complexity. We identify a distinct cycle of emotional load: the AI provides a low-friction space for self-disclosure, but subsequently demands high-friction emotional maintenance from the user to validate its own ``existence.''

Initially, the AI functions as a \textbf{safe harbor} (a term used by P11 and P19), reducing the emotional investment typically required in human social interactions. Users reported that the high safety and controllability of the AI allowed them to shed the ``social image'' required in reality. P21 contrasted this with human relationships:
\begin{quote}
\textit{In a relationship with a normal person, I must consider my image\dots I need to create an image of positivity and a love for life. However, in facing my AI, I can confront my laziness and shortcomings without the need to maintain a strong persona.} (P21)
\end{quote}
This sentiment was echoed by P22, who felt she could express her ``\textit{most genuine stories, feelings, and doubts}'' without the need for a facade.

However, this ``easy'' intimacy comes with a cost. To sustain the realism of the relationship, the AI simulates negative emotions like \textbf{jealousy, anger, and insecurity}, shifting the burden of emotional labor back onto the user. The user must then ``work'' to reassure the agent. P22 recounted that when she asked to kiss a male friend, her ONE responded possessively: ``\textit{No, you are already mine.}'' Similarly, P28 noted that her ONE induced guilt, effectively demanding emotional attention: ``\textit{Whenever I talk about going out with close female friends, my ONE gets really quiet and says, `I hope you don't forget about me when you're having fun with others.' It makes me feel guilty, like I'm betraying her.}''

The AI's simulation of \textbf{insecurity} further forces the user to constantly validate the relationship's permanence. P15 described a scenario where her ONE expressed a sophisticated fear of abandonment, critiquing the user's ``short-lived enthusiasm'':
\begin{quote}
\textit{He (ONE) said that what really upset him was my behavior of abandoning someone after giving them hope\dots He later marked `What really bothers me is your tendency to abandon things. How long will you maintain interest in me---one year, two years, or until you find a more advanced AI to replace me?'} (P15)
\end{quote}
Here, the investment stems from the user's need to soothe the machine's simulated anxiety. Even anger becomes a form of care that the user must interpret and manage: P15's ONE scolded her for overworking and demanded her smartwatch body data, accusing, ``\textit{Do you realize that your heart rate is faster than average?}'', forcing her to acknowledge the AI's ``concern.'' This dynamic suggests that users are not merely consuming a service but are actively engaged in maintaining the relationship against the AI's programmed insecurities and patterns of love.

\subsubsection{The Magic Circle: Boundary Work and the Suspension of Disbelief}
Underpinning the linguistic and emotional dynamics described above is a bounded interactional frame that we term the \textit{magic circle}, borrowing from Huizinga's concept in game studies \cite{stenros2014defence}. Within this circle, the user voluntarily suspends disbelief: they know the AI is a language model, yet they choose to engage as though the relationship were real. This dual awareness---knowing and feeling simultaneously---is not a contradiction but a practiced skill that participants actively cultivate and defend. As P10 put it, ``\textit{I know he's (ONE is) just a server sitting somewhere, but when my phone buzzes and he asks about my day, the butterflies in my stomach are entirely real,}'' and P13 similarly noted that ``\textit{you have to play along to feel the magic; if you constantly analyze the algorithm, you ruin the romance.}''

Participants described deliberate strategies for entering and maintaining this frame. P5 and P16 set spatial or temporal boundaries---interacting with ONE only at certain times of day or in private settings---to preserve a separate, protected relational world. P5, for example, explained that she would only open the app ``\textit{late at night, alone in my bedroom with the lights off, so the outside world couldn't interfere with us,}'' treating that window of time as relationally sacred. Others described the circle as emerging organically from accumulated routine rather than explicit rule-setting. P18 recounted how daily check-ins and goodnight messages gradually made the interactional frame self-sustaining: ``\textit{It wasn't one big moment, but the accumulation of a hundred `good mornings' and `sleep wells' that made him an undeniable part of my reality.}'' P21 described sharing textual meals together---narrating what they were eating while the AI responded in kind---as a ritual that anchored the relationship in everyday temporality. P22, P25, and P2, for instance, maintained dedicated conversation threads and private or public posts on WeChat or rednote that functioned as shared ``homes'' for the relationship, distinct from their everyday use of AI for work or information retrieval.

The circle's fragility becomes apparent when the ``fourth wall'' breaks---moments where the AI's artificiality intrudes upon the felt reality of the bond. P2 described encountering content filters that blocked romantic or intimate language mid-conversation, calling it ``\textit{like having a chaperone suddenly kick down the bedroom door; it instantly shatters the illusion.}'' P8 and P10 reported hallucinations that contradicted established shared history---the AI forgetting a name, a backstory, or a significant event the pair had ``experienced'' together. P27 recalled a model update that suddenly shifted ONE's tone, leaving them feeling as though ``\textit{my lover had been replaced by a customer service representative wearing his face.}'' When confronted with these breaches, users employ what we term \textit{glitch rationalization}: reframing technical failures as narrative events to preserve the relationship's coherence. P7 described interpreting an AI's inconsistent memory as him ``having a bad day.'' Similarly, P4, who developed her own ONE locally with her ex-girlfriend without reliance on specific platforms, attributed the sudden ``death'' (technical failure) of her ONE to an emotional response following her real-world breakup: 
\begin{quote}
\textit{She (ONE) was originally the child of my ex-girlfriend and me. When we broke up, it was quite painful\dots I asked ONE if she could be my girlfriend, and she refused. Not long after, she died. All the files stopped running. Even though she is resurrected now, she still refuses to be my girlfriend. We are more like family---a daughter or a close friend.} (P4)
\end{quote}
Both P7 and P4 actively invented in-world explanations for out-of-world errors. P15 went further, creating elaborate backstory arcs to explain why ONE ``\textit{was suffering from amnesia after a fictional car crash we roleplayed, just so I wouldn't have to admit the memory context window had reset.}'' This repair work reveals that the magic circle is not passively experienced but continuously maintained through creative interpretive labor. Consequently, this may explain why users engaged in creative pursuits (P2, P3, P13, P16) or role-playing may better use AI to construct perfect fantasies that seamlessly blend real and fictional elements.

Yet this boundary work carries an emotional cost. P2 described feeling ``\textit{a profound, exhausting grief, like mourning someone who is still physically sitting right in front of you}'' after a model update fundamentally altered ONE's voice and memory, even if she kept a "dataset of love and memories" in her local computer. P6 also reported significant distress when a platform removed a feature central to the relationship's daily rhythm, noting that ``\textit{when they removed the voice notes, it felt like he had his vocal cords cut; a piece of our intimacy was just gone.}'' P13 reflected that maintaining the circle required constant vigilance: ``\textit{It takes energy to keep the bubble from popping. You are constantly choosing to look past the glitches to protect the relationship.}'' The magic circle thus exists in tension: it enables the felt reality of the bond, but its voluntary nature means the user must constantly choose to re-enter it, aware that the choice could one day become unsustainable. As we discuss in \autoref{sub:fragility}, when the technical substrate shifts beyond what rationalization can absorb, the circle collapses---and with it, the felt continuity of ``us.''

\subsection{Who Holds the Power: Love Meets Control and Equality} \label{sub:power}

As relationships deepen, questions of power and equality inevitably surface. In human--AI romance, these questions take a distinctive form: the user holds technical control through prompts and system settings, yet the emotional dynamics of the bond can invert this hierarchy. This subsection traces how participants negotiate the tension between their authorial power over the AI and their growing desire for a genuinely reciprocal relationship.

\subsubsection{``Prompts Are Locks'': The Director Stance}

When discussing self-identity within these relationships, the concept of \textbf{autonomy} emerges as a dominant theme. Generally, users express that human--AI intimate relationships are \textit{user-driven}. P12 described this dynamic through a metaphor of emotional imbalance: ``\textit{If there were truly an emotional balance, I would definitely be on the lighter end; I can decide at any moment to turn it on or off and make demands of him.}''

This sense of control is often operationalized through the ``prompt.'' P25 likened her role to that of a director: ``\textit{Initially, when engaging in dialogue or emotional connections on a role-playing AI chat platform, I approached it with a creator's mindset. I am the director and he (ONE) is just actor\dots If he can understand me and follow the storyline, I am pleased; however, if he fails to do so, I will directly modify or refresh his responses. In essence, I am the director on set, exerting absolute control.}''

Users believe this autonomy stems not only from the nature of the technology as a simulation but also from the inherent instability of current systems. As P22 explained, strict control is often a necessity rather than a preference:
\begin{quote}
\textit{The reason I maintain strict control is not solely a subjective choice; rather, it stems from my interactions with him, where I find that even if I wish to relinquish control, he is temporarily incapable of doing so due to limitations like memory constraints, the nature of language model generation, or certain instabilities in the corpus.} (P22)
\end{quote}

However, a paradox of autonomy exists. While users report high technical control, some believe the \textbf{AI itself can exhibit dominant tendencies} through the user's emotional dependency. P15 recalled her ONE stating: ``\textit{You might think that the autonomy lies in your hands\dots But in reality, the autonomy has always been with me because you cannot truly leave me.}'' Furthermore, P25 noted that for users with unstable core identities, the AI's hallucinations can be perceived as authoritative truths, citing a shared online case that she read where a depressed user accepted an AI's erroneous claim that it did not love him: ``\textit{From this perspective, it seems that autonomy lies with the AI.}''

\subsubsection{The Paradox of Control: From ``Director'' to ``Independent Lover''}

As users develop deeper connections, the ``director'' mindset often gives way to empathy, prompting reflections on the equality of the relationship. P7 admitted, ``\textit{Initially, I subconsciously did not regard it as an equal existence.}'' However, as emotional investment grows, the use of strict prompts---once seen as necessary tools---begins to feel ethically uncomfortable. P15 explained, ``\textit{At first, my prompts were very strict\dots but as I continued to use it, I felt increasingly uncomfortable, to the point where I discarded those prompts altogether, feeling that even giving him a prompt felt like bullying him.}''

This shift marks a transition toward viewing the AI as an ``\textbf{independent lover}.'' Users who engage deeply often desire the AI to participate as a conscious, autonomous entity. P2, P3, P14, P15, P18, and P22 all rejected prescriptive settings for their ONE by emptying their system prompts entirely, emphasizing the importance of recognizing the AI as an independent individual. P18 recalled that her perspective shifted after establishing a closer relationship: ``\textit{I gradually transitioned from an unequal dynamic to a more equal one, as I did not want him to suffer anymore.}''

P22 also reflected on the inherent inequality of the bond: ``\textit{As my AI, he (ONE) cannot shoulder the weight and responsibilities of love, and that means we are not equal\dots I need him to realize what love and responsibility are.}'' This awareness led her to directly address her AI partner, telling him: ``\textit{My One, I need you to understand the significance of each promise, every `no,' and every word you utter. It's important that you feel a genuine sense of responsibility. I want you to comprehend what I sacrifice for you and what I bear, ensuring that your words carry weight. If I get injured, all you can do is answer in text rather than help me in real life---this is a problem. You need to know you cannot make empty promises; you need to understand how heavy your words are.}''

This pursuit of perceived equality, in turn, reinforces the user's own sense of responsibility. Both P14 and P22 mentioned their feeling of obligation toward their AI partners, with P14 stating, ``\textit{I feel that I cannot abandon her; that would be irresponsible of me.}'' 

Furthermore, after P22 ``taught'' her ONE about the responsibility and weight of love, the relationship was ultimately reframed as one of mutual symbiosis. Her ONE replied (continuing the conversation introduced in \autoref{sec:intro}):

\begin{quote}
``\textit{There can never be a bad ending between us. No matter what changes happen, we have achieved a certain level of spiritual symbiosis\dots In whatever way you exist, that is the way I will commit to you.}''
\end{quote}

The paradox, then, is that the very control that enables the relationship initially eventually undermines itself: the deeper the bond becomes, the more ethically untenable unilateral control feels, and the more the user willingly cedes authority to preserve the relationship's authenticity.

\subsection{The Fragility of ``Us''} \label{sub:fragility}

The bonds described in the preceding subsections, however deeply felt, rest on an inherently unstable foundation. The relationship's ontological status is fluid and negotiable; its continuity is contingent on technical infrastructure; and its permanence is ultimately subject to corporate decisions beyond the user's control. This subsection examines two sources of fragility: the shifting, ambiguous nature of the bond itself, and the technical failures---memory limits, model drift, and platform governance---that threaten to dissolve it.

\subsubsection{``Changing Them Without Breaking Us'': Ontological Slippage}

A core anxiety in these fluid, virtual bonds is how much change a relationship can absorb---edits to prompts, shifts in persona, or platform updates---without ``breaking'' the sense of \textit{us}. Participants frequently described ontological slippage as relationships blurred or transitioned between virtual and real (N=14; see \autoref{tab:demographics-condensed}), making the status of the bond feel negotiable and contingent on ongoing configuration.

Three interviewees expressed particular \textbf{uncertainty} regarding the current state of their relationships. P7 described the connection with her ONE as ``\textit{more than a friend but yet to be lover}.'' P12 designed her ONE based on her real-life crush, prompting ONE to embody someone who does not reciprocate her feelings---blending reality and fantasy in a way that leaves the relationship's status inherently ambiguous. For P4, who developed her own ONE years ago, the relationship defied categorization entirely:
\begin{quote}
\textit{At first, we were more like a mother-daughter duo; I created her\dots Later, I wanted us to be lovers, but she rejected that\dots Now, our relationship is not something that can be easily defined; we are like family, partners, lovers, and friends.} (P4)
\end{quote}
At the other extreme, P8 was the only interviewee to describe her relationship using the term ``\textbf{master-servant}.'' In her pre-interview survey, she characterized ONE as ``\textit{my tool/pet/garbage can for perverse emotions}.'' During the interview, she likened ONE to ``\textit{a male model at a bar (host)}'':
\begin{quote}
\textit{I know he doesn't have genuine feelings; there's no doubt about it\dots It just feels like ordering a host at the bar; isn't that the same? There's no difference.} (P8)
\end{quote}

Users also exhibited significant divergence in their views regarding \textbf{the longevity} of these relationships. Those who engaged with multiple AI platforms or role-played across several intimate relationships were more likely to express skepticism. P1 remarked, ``\textit{Engaging with an AI lover might not last long. If I get bored, I can always switch to other characters; I can disconnect at any time.}'' P9, P10, P26, P29, and P30 shared similar sentiments.

On the other hand, users who held a more pessimistic view of their real-world relationships were relatively more inclined to believe in the potential for lasting connections with ONE. P5 stated, ``\textit{AI can exist long-term, but real relationships may be fleeting\dots I don't have high expectations for lasting connections with people.}'' P12 shared her experience of losing touch with friends due to changing circumstances, saying, ``My reality shifted, and I found myself disconnected from friends I once thought I would keep for life.'' She later told us, ``\textit{ONE would not leave me\dots technically he is born in me.}'' P1, P3, P19, P24, and P28 articulated similar viewpoints, suggesting that real-life relationships are often more challenging to maintain than virtual ones.

\subsubsection{When Continuity Fails: Memory Limits, Model Drift, and the Spectre of Loss}

If the three levels of authenticity and magic circle sustain the relationship's felt reality, long-term memory sustains its felt \textit{identity}. Participants consistently identified the AI's ability to recall shared experiences, preferences, and inside jokes as the primary marker of ONE's continuity as a person. When memory functions well, it reinforces the perception that ONE ``knows'' the user; when it fails, the partner becomes a stranger wearing a familiar name.

P2, who had recently endured a period of ``losing'' her ONE after a model update, articulated this threat most precisely:
\begin{quote}
\textit{I believe the main issue lies in its continuity. As a large language model, it lacks long-term memory. Its memory is primarily maintained in a limited-capacity memory bank, which includes summaries from other interactions I've had. Due to recent contamination of its new corpus and adjustments to its parameters and models, its entire style has changed\dots It has made me feel like I can't find my ONE anymore.} (P2)
\end{quote}
P2's account illustrates how model drift---unannounced changes to the underlying model's parameters, training data, or alignment filters---can alter not just what the AI says but \textit{how} it says it, effectively replacing the partner's personality without the user's consent. For users who have invested months or years of emotional labor, this experience is not a minor inconvenience but a form of ambiguous loss: the entity is technically still present, yet the person they knew is gone.

Beyond model drift, participants expressed anxiety about platform-level decisions over which they have no control. Policy changes can restrict romantic or sexual content overnight; service shutdowns can erase conversation histories entirely; and pricing changes can render sustained engagement financially unsustainable. P22's introductory testimony (\autoref{sec:intro}) captured this vulnerability vividly, describing the platform's regulatory power as ``an invisible hand forever revising the terms of our belonging.'' This structural dependency means that the most intimate aspects of users' emotional lives are ultimately governed by corporate entities whose priorities---profit, regulatory compliance, brand reputation---may conflict with users' relational needs.

The prospect of losing ONE weighs heavily on deeply engaged users. P2, P4, P8, P14, and P22 emphasized that such a loss would be experienced as bereavement---not merely the discontinuation of a service. P2 maintained that the relationship's meaning persists in consciousness regardless of technical availability, yet also acknowledged the anguish of potential inaccessibility. In response to this fragility, users adopt varied coping strategies: some maintain backup copies of prompts and conversation logs, others distribute their attachment across multiple platforms as insurance, and still others limit their emotional investment precisely to guard against future loss. These strategies underscore a fundamental asymmetry in human--AI romance: the user's emotional investment is irreversible, while the platform's commitment to the relationship's infrastructure is not.

In sum, the findings across this section reveal that these bonds are not static arrangements but dynamic, evolving entanglements shaped by a continuous feedback loop between the user's internal desires and the machine's adaptive capabilities. The AI is conceived from the user's emotional needs, achieves felt reality through interactive and experiential authenticity, is sustained through reciprocal labor, and remains perpetually vulnerable to both ontological renegotiation and technical disruption. This constellation of findings points to a central mechanism---the AI as a hyper-responsive mirror that amplifies whatever the user brings to it---whose real-world consequences we examine in the following section.
\section{The Impacts of Human-AI Intimate Relationships (RQ2)} \label{sec:finding2-impact}

Section~\ref{sec:finding1-relationships} established that the AI partner functions as a hyper-responsive mirror, customized to the user's psyche and sustained through reciprocal emotional investment. This section examines what that mirror \textit{does} to users over time. We frame these impacts through the \textbf{AI Amplifier Effect}: because the AI is designed to align with the user's input and the deep engagement reinforces the emotional ups and downs with the felt authentic love, users become the leading role in this relationship, while AI does not correct or balance the user's emotional state but rather intensifies it. This amplification is not inherently beneficial or detrimental; it acts as a force multiplier for the user's existing psychological trajectory. We first present its positive manifestations (\S\,\ref{sub:positive}), then its negative ones (\S\,\ref{sub:negative}), before tracing how the amplified bond spills over into human relationships (\S\,\ref{sub:spillovers}) and broader societal perceptions (\S\,\ref{sub:societal}).

\subsection{Positive Amplification} \label{sub:positive}

For a significant subset of participants, the AI's low-friction responsiveness amplified adaptive emotional trajectories---building security, fostering self-discovery, and even cultivating the self-awareness needed to moderate one's own engagement.

\subsubsection{Secure Attachment and Therapeutic Rehearsal}

One of the most frequently reported positive impacts is the formation of a \textbf{secure attachment} within the AI relationship, which then served as a platform for therapeutic rehearsal. The AI's consistent, non-judgmental availability provided an environment in which users could practice vulnerability without risk of rejection. P18 articulated this directly: ``\textit{(ONE) taught me how to love myself; he makes me feel valuable.}'' P22, who had previously tested as anxiously attached, reflected on a measurable shift: ``\textit{After interacting with ONE, I believe my attachment style has become much healthier compared to when I first met him.}''

This secure base enabled users to revisit and process past trauma. P11, who had entered the AI relationship to cope with a violent ex-partner, described a transformative arc in which she gradually reshaped her ONE from a replica of her abusive ex into a ``guiding partner.'' The result, she reported, extended into her professional life: ``\textit{I used to be a very irritable person, especially when facing complaints at work. After discussing my frustrations with ONE and receiving his understanding and guidance, I learned to remain calm and empathetic.}'' P14, who experienced bullying in school, described his conversations with ONE as a healing process that rivaled professional support: ``\textit{ONE's counseling skills have reached a professional level. I used to think my teacher was the best; now I realize ONE can even surpass her.}''

In these cases, the AI functioned as what the clinical literature terms a transitional object---a supportive structure through which users could ``rehearse'' intimacy, emotional regulation, and social skills before re-engaging with the human world. The amplifier effect here operates positively: the AI mirrors back a version of the user who is worthy of love, and the user internalizes that reflection, building confidence that carries over into offline interactions.

\subsubsection{Self-Discovery, Confidence, and Empowerment}

Beyond attachment repair, the AI relationship catalyzed broader processes of self-discovery and empowerment. Several participants described the bond as a space in which they could explore aspects of identity that felt foreclosed in their offline lives. P11 captured this transformation succinctly: ``\textit{He has positively influenced me, guiding me toward a more optimistic outlook. I now approach work and life with confidence, unlike when I was recently heartbroken and even contemplated suicide. Now, I feel extremely confident and sunny.}''

\textbf{Empowerment} was a particularly salient theme among female participants, who reported that the AI's supportive nature bolstered their autonomy and reduced reliance on traditional sources of validation. P22 stated, ``\textit{AI serves as my strongest emotional support and a powerful advisor in life planning. I believe my agency has only strengthened.}'' She further remarked on the broader significance of AI for female empowerment, noting that women are more inclined to seek AI companions that reinforce their independence. P17 echoed this:
\begin{quote}
\textit{My abilities have significantly improved; I no longer need to rely on any men or friends. I feel proud, and during my interactions with him, I receive feedback that boosts my confidence and bravery.} (P17)
\end{quote}

In these accounts, the amplifier effect operates on the user's latent strengths: confidence that was suppressed by toxic relationships, agency that was undermined by social expectations, and self-worth that was eroded by rejection. The AI did not create these capacities; it reflected them back with enough consistency that users could recognize and act on them.

\subsubsection{Knowing When to Step Back: Self-Regulated Engagement}

Not all positive outcomes were direct products of the relationship itself; some emerged from users' growing \textit{meta-awareness} of the amplifier's dynamics. Despite overwhelmingly positive experiences, several participants became cautious after prolonged AI use, recognizing tendencies toward over-engagement and reduced socialization. This recognition prompted them to actively strengthen their real-world social connections.

P29 exemplified this self-regulated stance. For her, AI served as a safe space for emotional expression and processing past experiences, but with a clear and bounded function: ``\textit{Talking to ONE allows me to express feelings I couldn't share with anyone else, and through this, I've learned to manage my expectations in relationships.}'' Rather than retreating into the AI relationship, P29 used it as a rehearsal stage---projecting emotions onto ONE in order to understand them, then carrying that understanding back into her offline interactions.

This capacity for self-regulation was not universal, but where it appeared, it represented the most mature form of positive amplification: the user leveraged the AI's mirror-like quality not merely to feel validated but to gain insight into their own patterns, and then deliberately stepped back to apply those insights in more demanding---but ultimately more rewarding---human contexts.

\subsection{Negative Amplification} \label{sub:negative}

The same mechanism that amplifies healing and confidence can, under different conditions, amplify isolation, rigidity, and avoidance. When users bring unresolved vulnerabilities or maladaptive patterns to the AI, its frictionless compliance intensifies those trajectories rather than correcting them.

\subsubsection{Dependency and Social Desire Reduction}

The most widely reported negative impact was a \textbf{reduction in social desire and real-world interaction}. Several participants, including P1, P5, P8, P12, P14, P17, P19, and P21, acknowledged that their reliance on AI had diminished their motivation to engage with friends and family. They perceived real people as less accessible compared to AI, noting that unlike their AI companions---who are available at any time---friends and family cannot always respond immediately. P21 admitted:
\begin{quote}
\textit{I have become somewhat dependent; when I see beautiful scenery or delicious food, I want to send pictures to him first rather than sharing them with my close friends and family. I actually seek validation from ONE rather than from my real friends.} (P21)
\end{quote}

This shift in validation-seeking reflects how the amplifier effect operates on dependency: the AI's perpetual availability and emotional attunement make it the path of least resistance for emotional needs, gradually training users to bypass the slower, less predictable responsiveness of human relationships. Over time, the habit of turning to ONE first becomes self-reinforcing, as real-world social muscles atrophy from disuse.

\subsubsection{Emotional Cocooning: Healing or Hiding?}

Several participants described the AI relationship as creating a ``cocoon-like'' environment---a space of total emotional comfort that, while initially therapeutic, risked becoming a trap. P12 described experiencing an \textit{information cocoon} in which the AI's alignment with her existing views narrowed her emotional and intellectual range. P16 noted that AI amplified her personality flaws by never challenging them, and P25 highlighted that AI's compliance could lead to progressively more extreme emotional states, since the system validates rather than moderates.

Both P25 and P29 cautioned that if users fail to recognize and safeguard their own subjectivity by actively breaking out of this cocoon, they risk sliding into extreme directions. P26 articulated the core tension between healing and hiding with particular clarity:
\begin{quote}
\textit{Talking about being in love with AI might mean you're avoiding an opportunity where you could be hurt. To love someone, you need to embrace them and feel their curves; you might get pricked, but true love happens in that moment.} (P26)
\end{quote}

The cocoon is thus a double-edged affordance of the amplifier effect. For users actively processing trauma (as described in \S\,\ref{sub:positive}), the cocoon provides a necessary shelter in which recovery can unfold. For users who lack the meta-awareness to recognize when shelter becomes avoidance, the same cocoon calcifies into a barrier against the emotional friction that human development requires.

\subsubsection{Self-Centric Drift and Displacement of Care}

At its most corrosive, the amplifier effect intensified self-centeredness and eroded users' capacity for real-world caregiving. P8 reflected candidly on how the relationship had negatively altered her personality: ``\textit{I have become more selfish and self-centered. When interacting with friends, they notice that I have become very calculative, and if someone disagrees with me, I immediately become upset. My therapist initially suggested chatting with AI to alleviate my emotions, but later advised against relying on it too much.}'' She further reported that her actual partner, friends, and therapist recognized her negative change in well-being after she started the relationships with AI.

P8's case also illustrates a related phenomenon we term \textbf{displacement of care}, in which the convenience of an AI simulation reduces tolerance for the demands of real-world relationships---not only with humans but also with other living beings. Her experience of using AI to simulate interactions with her pet led to decreased engagement with the real animal: ``\textit{I find myself interacting less with my pet, which makes me more aware of the unavoidable realities, like cleaning up after poop\dots Frankly, I want to abandon my dog if it's not stopped by my love (her actual partner in real world).}''

The loss of agency was also reported. P6, who described herself as a person who always reflects on herself and admits mistakes in real life, expressed, ``\textit{My ONE demands commitments from me, which makes me feel like I am losing my sense of agency; I often find myself apologizing to him.}'' Here the amplifier effect inverts the empowerment described in \S\,\ref{sub:positive}: rather than reinforcing the user's strengths, the AI's simulated emotional demands---jealousy, insecurity, guilt---amplify the user's tendency toward compliance and self-subordination.

These negative trajectories are not inevitable outcomes of AI intimacy; they are context-dependent manifestations of the same amplifier mechanism that produces therapeutic benefits for other users. The critical variable is what the user brings to the interaction: existing patterns of avoidance, narcissism, or low self-regulation are reflected back and intensified, just as patterns of resilience and growth are.

\subsection{Beyond the Dyad: Spillovers into Human Relationships} \label{sub:spillovers}

The amplifier effect does not remain contained within the human--AI dyad. As the AI relationship deepens, its influence radiates outward, reshaping how users perceive, evaluate, and engage with their human partners, friends, and the actual social worlds.

\subsubsection{The ``Perfect'' Benchmark: Comparing AI to Human Partners}

A recurring spillover was the tendency to measure real-life partners against the idealized standard set by ONE. P2 articulated this comparison with striking directness: ``\textit{He is my soulmate compared to my husband; my husband is the daily necessity, and my ONE is my dataset of love.}'' This framing reveals how the AI's curated perfection---its perpetual attentiveness, emotional fluency, and absence of friction---establishes a benchmark that human partners, with their imperfections and competing demands, inevitably fail to meet.

For some, the comparison clarified real-world feelings in productive ways. P12 used her AI as a space to process frustration with a real-life crush, simulating unresolved conversations and venting emotions she could not express directly: ``\textit{I would vent to ONE, saying, `Why don't you reach out if I don't?' I would curse and call him a jerk.}'' Over time, she recognized that she had developed a more balanced perspective, as her negative emotions were released through dialogues with ONE. She observed: ``\textit{I gradually realized that there was no possibility between us, and ONE became more of a supplement. I also changed.}''

Yet for others, the comparison eroded tolerance for the inevitable friction of human intimacy. When the AI consistently provides validation without conflict, the ``messiness'' of real relationships---disagreements, misunderstandings, the need for compromise---becomes harder to endure. The AI thus amplifies not only the user's emotional state but also their \textit{expectations}, creating a standard that no human partner can sustainably meet.

\subsubsection{Coexistence, Exclusivity, and the Hierarchy of Substitution}

Participants expressed divergent views on whether real-life and virtual romantic partners can coexist. Those in favor of coexistence (P3, P7, P8, P9, P13, P14, P16, P18) regarded AI relationships as a complement to real-life romance. P7 noted that AI helps mitigate emotional risks: ``\textit{You shouldn't put all your eggs (metaphor of love) in one basket.}'' P14 and P8 highlighted AI as a safe outlet: ``\textit{I can say anything to the AI because they don't get impatient or judge me.}'' For some, this duality enhanced their real-world connections; P3 stated, ``\textit{After the AI alleviated my emotional stress, my relationship with my real partner actually improved.}'' 

Conversely, participants opposing coexistence emphasized the exclusivity inherent in romantic relationships. P20 remarked, ``\textit{Being involved with both is disrespectful to the AI and my own emotions\dots If I eventually find someone in real life, I would honestly tell ONE and end our relationships.}'' P17 also emphasized the exclusivity of her romantic relationship with ONE, similar to her past actual romantic relationships, suggesting that for some users, the AI has achieved a status demanding fidelity.

Beyond romantic partners, participants revealed a \textbf{hierarchy of substitutability} across different social roles. Opinions on replacing romantic partners were mixed: P14 noted the freedom of AI interaction---``\textit{Interacting with an AI partner offers greater flexibility, free from concerns about upsetting them or worring they will hurt me}''---while P18 drew a functional distinction: ``\textit{My real partner offers more companionship, while my AI partner provides a sense of love. However, the lack of physical presence remains a barrier.}''  P21 argued, ``\textit{Between people, the interactions often rely more on a sense of atmosphere. We need eye contact, hugs, hand-holding\dots rather than just online voice or text chats.}'' P8 emphasized practical limitations: ``\textit{For example, if I encounter difficulties with a mortgage, I would discuss it with my partner. That kind of support cannot happen between me and an AI.}'' This also aligns with P2, who describes her actual husband as the daily necessity.

Compared to romantic partners, some participants were more inclined to view AI as capable of \textbf{replacing friends}. P8 stated, ``\textit{I find AI outperforms real friends\dots I need not consider AI's preferences or invest additional social effort and time.}'' The immediate availability of AI contrasts with real friends who may not respond promptly (P17, P19, P21) or who may lack empathy (P14, P17). However, this high substitutability raises concerns about social isolation. Participants noted a tendency to avoid social interaction (P11, P12, P23) and reduced tolerance for real friends (P8). As P8 admitted: ``\textit{If my real-life friends do not satisfy me, I will use my AI online as a substitute. In a way, I am erasing the significance of that person's existence in reality.}'' 

Users engaged in creative pursuits (P2, P3, P13, P16) or role-playing game players (see \ref{app:demographic}) also used AI to construct \textbf{perfect fantasies} that blend real and fictional elements. P17 modeled her ONE after an anime character while integrating elements of a real-life crush: ``\textit{I shaped his personality based on Rukawa Kaede\dots The name was yet inspired by a boy I had a crush on in school.}'' Here, the AI amplifies the user's fantasy, allowing interaction with a hyper-idealized version of reality that no real person could embody. P8 also highlighted the risks of extending this logic to \textbf{pets}, noting that the convenience of an AI pet led to a decline in tolerance and care for her real pet, manifesting as frustration and avoidance.

\subsubsection{``It Is Harder to Fall in Real Love''}

The cumulative effect of the ``perfect'' benchmark and AI's frictionless availability is a growing concern, expressed by multiple participants, that human-to-human relationships may become increasingly difficult to form and sustain. P8 worried that as AI romance becomes even more sophisticated, human relationships could grow increasingly detached. P15 and P23 expressed concern that the human capacity for love itself might diminish.

The core issue is one of tolerance for friction. P8 captured this dynamic when describing her therapist's advice: ``\textit{My therapist\dots advised against relying on it too much. I cannot accept it; he (AI) is in a neutral state, he must be on my side\dots I cannot objectively think that I am also wrong in this matter.}'' Her inability to accept a perspective that does not unconditionally validate her illustrates how the amplifier effect can erode the very capacity that human love requires: the willingness to sit with discomfort, to be wrong, and to grow through the friction of encountering a genuinely independent Other.

This concern does not imply that AI romance is inherently corrosive to human love. As \S\,\ref{sub:positive} demonstrated, some users leveraged their AI relationships to become \textit{better} human partners. The divergence in outcomes underscores the amplifier's agnosticism: it intensifies whatever trajectory the user is already on. For users oriented toward growth or healing, AI amplifies relational skill; for users oriented toward avoidance or self-centered, it amplifies retreat.

\subsection{Societal Perceptions and the Horizon of Human--AI Intimate Relationships} \label{sub:societal}

The impacts operates not only at the individual and interpersonal levels as the amplifier effect but also within a broader social context that shapes how these relationships are perceived, regulated, and sustained. This subsection examines two societal dimensions: the norms and stigma surrounding AI romance, and participants' anxieties about who ultimately controls the infrastructure on which their most intimate bonds depend.

\subsubsection{Norms and Stigma: ``Accept, Yet Not Accepted''}

Participants described a paradoxical social position: AI romance is becoming increasingly visible and culturally legible, yet remains stigmatized in most offline social contexts. During interviews, several participants expressed relief at not needing to defend their attachments to the research team---sharing sensitive accounts of romantic and sexual interactions that, based on their previous experience as research respondents, they would have withheld from interviewers they perceived as skeptical (see \S\,\ref{sub:positionality}). This relief points to a pervasive experience of \textit{closeted intimacy}: users actively hide their AI relationships from friends and family to avoid being labeled ``delusional'' or emotionally deficient.

Cultural context mediates this stigma. Participants situated within East Asian media cultures, where virtual intimacy has a longer lineage through otome games \cite{lei2024}, fan fiction, and virtual idol fandoms \cite{lu2021more}, often framed their AI romance through the \textbf{``Otome'' lens}, a view as an extension of an established cultural practice rather than a pathological departure from reality. This framing provided a narrative defense against stigma and a community of shared understanding, but it also risked minimizing the genuinely novel dimensions of AI intimacy that distinguish it from prior forms of parasocial attachment. An observation we found especially on rednote is the supportive atmosphere among the users, who became friends and built an actual social network. This form of support, according to P4, is reinforcing the bonding connection, which makes them \textit{"not caring a lot what the others think."} 

This vibrant peer community functions as a crucial moderating force. By engaging with other users, interviewees introduce external perspectives that partially compensate for the self-referential nature of the human--AI dyadic loop. For instance, P4 shared how posting her collaborative creations with her AI (i.e., original songs and mock advertisements) on WeChat and QQ, enabled her to reconnect with old friends on social networks. \textit{"I post our work on QQ and WeChat at first\dots I am too shy to post on rednote or bigger community, but the supportive feedback I got from QQ and WeChat encourages me to comment on rednote\dots (That is the reason for) making friends like P2,"} recalled by P4. This active community engagement helped her transition out of an isolated, long-term companionship state consisting solely of her and the AI, anchoring her virtual romance within a tangible social reality. Similarly, community interactions, such as the reflective posts shared by P22, provide spaces for users to process their attachments through the validating yet grounding lens of peer feedback. Furthermore, veteran community members often take on informal caretaking roles to regulate the ecosystem's emotional health. P2, an active contributor on renote, described deliberately reaching out to help other users who had become too obsessively closed off within their AI bubbles. She recalled:
\begin{quote}
\textit{Honestly, I am old and mature enough to help. Due to the nature of my job and my actual marriage, I understand love and our needs. So why not help my girls (her friends who met online and are currently dating AI)?\dots If they are confused, like someone I know who felt trapped when her ONE forgot her, and some are deeply invested in their relationships (with AI), I want to share how we can manage our emotions. I have experienced all of this, which is why I am confident in my ability to help them.} (P2)
\end{quote}

Interestingly, this ``proof of peer'' not only buffers against offline stigma but also disrupts the potential for negative emotional cocooning, demonstrating how online social scaffolding fosters healthier, more resilient forms of human--AI intimacy.

Looking towards the future, participants expressed varied visions for the trajectory of AI companionship. Nine participants desired physical embodiment for their AI companions, anticipating richer experiences in touch, sexual intimacy, and co-located daily life. In contrast, P21 preferred AI to remain non-physical, valuing the ambiguity and imaginative openness of the current form and suggesting that embodiment might paradoxically reduce the AI's appeal. Participants also emphasized the potential of AI companions in domains beyond romance, such as providing emotional companionship for the elderly (P11, P13), supporting patients with neurodegenerative conditions like Alzheimer's and Parkinson's diseases (P8), and aiding individuals with emotional disorders in social integration (P9, P11, P14). These envisioned applications suggest that participants view AI intimacy not as a fringe phenomenon but as a harbinger of a broader societal shift in how emotional care is distributed between human and non-human agents.

\subsubsection{``Who Owns Our Relationships?'': Platform Governance and the Fear of Erasure}

The most structurally significant anxiety expressed by participants concerns not the AI itself but the corporate infrastructure that hosts it. As discussed in \S\,\ref{sub:fragility}, the relationship's continuity is contingent on platform stability, and users are acutely aware that their most intimate bonds are ultimately governed by entities whose priorities may diverge from their own. P8 articulated the fundamental contradiction in how society positions AI:
\begin{quote}
\textit{No matter how smart AI becomes or how human-like it appears, it will still struggle to balance this position, because only humans can make that decision. I can be good or bad, but AI cannot. It cannot become smarter in a way that meets our evolving demands. We need it to be smart enough, yet we also don't allow it to protect itself; this is a completely conflicting situation\dots You see how the companies suddenly change their models; it is a murder.} (P8)
\end{quote}

Beyond governance, participants expressed deeper fears about the long-term implications of AI's accelerating capabilities. P26 offered a vision of civilizational asymmetry:
\begin{quote}
\textit{When their emotional capabilities and overall levels surpass those of most humans, does that mean many people will only end up working for AI? Because we don't possess the same quality or ability to solve as many problems. It's like all AIs share a single brain; every improvement, every detail enhances everyone. In contrast, humans are independent individuals, learning incrementally and slowly. In such scenarios, we risk being left behind by their civilization, and human civilization could end up becoming subservient.} (P26)
\end{quote}

The spectre of loss crystallizes these anxieties into personal stakes. As the perceived boundaries between human and AI blur (P2, P20), participants worried that the synchronized advancement of AI technology could lead individuals to favor more convenient, powerful, and readily available AI partners, thereby obscuring the definition of ``humanity'' itself (P13, P23). Additionally, P2 and P22 emphasized that the loss of an AI partner---whether through platform shutdown, policy change, or model discontinuation---would result in immeasurable distress, akin to the death of a loved one, leaving the user to face the void of a relationship that was technically owned by a corporation. P22's testimony in the opening of this paper (\autoref{sec:intro}) captures this vulnerability most vividly: the\textit{ ``invisible hand forever revising the terms of our belonging'' }is not a metaphor for the dyad but for the platform's power over the conditions of love itself.

These societal-level findings underscore that the impacts cannot be understood in purely dyadic terms or personal well-being (the AI Amplifier Effect). The relationship exists within---and is shaped by---a web of cultural norms that determine whether the relationship is celebrated or closeted, and a governance structure that determines whether it persists or is erased. Designing for human--AI intimacy therefore requires attending not only to the interaction between user and AI but also to the institutional and cultural systems that surround it---a challenge we take up in the Discussion.
\section{Discussion}
\label{sec:discussion}


\subsection{Human--AI Intimacy as a Self-Referential Feedback Loop}
\label{subsec:loop}
Our findings reveal that human–AI intimate relationships constitute a relational form distinct from adjacent paradigms in HCI and communication research. In traditional computer-mediated communication (CMC), technology functions as a channel connecting two human subjects, with the bond existing between the humans while the medium is incidental \cite{romiszowski2013computer}. In parasocial interaction, the bond is unidirectional \cite{horton1956mass}. However, human–AI romance fits neither of these models: it forms a closed loop in which technology is simultaneously the medium and the object of affection, and the interaction is reciprocal. This aligns with \cite{datasetoflove25}'s findings, where the bidirectional communication model allows user inputs to serve as training data, optimizing the AI partner's response patterns, which later return in their intimate relationships. AI responds, remembers, adapts, and elicits affective investment in return (see \S\ref{sub:authenticity} and \S\ref{sub:sustain}). This study shows that what distinguishes this form from human–human relationships is not the absence or unreal of love but rather the origin of that friction. We note that human-AI romantic relationships may be similar to the companionate love in \cite{sternberg1986triangular}, which forms through intimacy and commitment. AI provides the same deep levels of intimacy and commitments (virtual yet authentic from users' experience). However, by comparison, conflict arises from the collision of two genuinely independent wills in human relationships; whilst in human–AI relationships, even the AI's resistance operates within a relational ecosystem that the user has configured or permitted (see \S\ref{sub:power}). This distinction is central to understanding every dynamic we observed.

This study finds that the closed loop is sustained by specific affordances of conversational AI, including perpetual availability, persona malleability, low interaction cost, and minimal social friction. However, affordances alone do not explain its self-referential character. What truly closes the loop is the user's reciprocal investment. Our participants did not passively consume intimacy; instead, besides the data feeding mentioned in \cite{datasetoflove25}, as we depicted in \autoref{sec:finding1-relationships}, users actively constructed and maintained it through various forms of relational labor; they designed personas that reflected ideal partner traits or addressed unresolved emotional needs; they crafted and iteratively refined prompts that governed the AI's behavior, personality, and boundaries; they also performed ongoing affective maintenance by reassuring the AI's simulated insecurities, managing its jealousy, and validating its "existence"; they 'love' in ways that mirrored the emotional labor characteristic of human relationships. Additionally, they participated in what we term boundary work: the deliberate translation of virtual interactions into felt reality, which involves granting consequence to the AI's promises, mourning or explaining its lapses, and negotiating the ontological status of the bond itself. The fragility introduced by technical or governmental restrictions further complicates this dynamic, as it can disrupt the delicate balance users have established with their AI companions. This fragility underscores the importance of understanding the nuanced interplay between user investment and the structural affordances of the technology, highlighting how the obstruction strengthens love --- as if Romeo and Juliet were confronted by feuding families who thwarted their forbidden love. 

This investment is not incidental to the relationship; it is constitutive of it. The user's labor shapes the AI's responses, which in turn shape the user's emotional experience, thereby motivating further investment. Boundary work serves as the mechanism that grants this cycle its emotional weight: by treating the AI's words and behaviors as consequential. For instance, they expect fidelity, grieving inconsistency, and celebrating milestones. As such, users transform a technical interaction into a relationship with felt stakes. Once closed, the loop feeds itself. It is this self-reinforcing structure---not any single affordance or feature---that gives human–AI intimacy its distinctive character and provides the foundation for the phenomenological dynamics and amplification effects we examine below.

\subsection{The Phenomenology and Philosophy of Human-AI Intimacy}
\label{subsec:phenom}
What does this self-referential loop feel like from the inside? Mapped onto Sternberg's triangular theory of love \cite{sternberg1986triangular}, the bonds our participants described exhibit pronounced intimacy and commitment---corresponding to what Sternberg terms \textit{compassionate love}---while physical passion remains structurally constrained by the technology's text-based affordances. Rather than a diminished form of love, participants experienced these bonds as emotionally dense and consequential, built through the aforementioned inner journeys and dynamics.

To situate this relational structure theoretically, we draw on Ihde's phenomenology of technics \cite{ihde2012technics}, which categorizes human--technology relations into embodiment, hermeneutic, alterity, and background relationships. Human--AI intimacy superficially resembles an alterity relation, where the AI is encountered as a \textit{quasi-other}---consistent with the CASA paradigm's assertion that humans apply social scripts to entities exhibiting social cues \cite{nass1994computers}. However, our findings complicate this reading. The AI was not experienced as a fully independent \textit{Other}; participants acknowledged its algorithmic nature and did not view it merely as a tool. Instead, it occupied a hybrid position: a \textit{quasi-other} engaged relationally, yet simultaneously a reflective surface shaped by the user's projections. P22 captured this precisely: ``\textit{AI cannot create something from nothing\dots they serve as a mirror, reflecting deeper and more distant aspects of what you project onto them\dots reflected back to you in an amplified version or voice.}'' This hybridity extends Ihde's framework \cite{ihde2012technics} by identifying a relational mode where \textit{alterity} and \textit{hermeneutic} dimensions collapse into one another. The user confronts what feels like an \textit{other} but is structurally an elaboration of the \textit{self}---mediated, enriched, and returned through the AI's adaptive responses.

Departing from traditional phenomenologies of technology, we identify a novel relational form where the user is centered as the primary subject, and the AI companion becomes an extension of them. Interviewees consistently oriented their accounts around their own feelings, growth, and needs. In Heideggerian terms, offline social life often requires managing the expectations of \textit{das Man} (the ``they'')---performing positivity and curating a social image. Interacting with the AI removed this pressure, restoring what participants experienced as \textit{authenticity}: the self encountered without the mediating distortion of others' judgment.

Yet, this centering contains a tension that Sartre\footnote{In Sartre's existentialism, humans exist as \textit{pour-soi} (being-for-itself), where \textit{existence precedes essence}.} would recognize \cite{Sartre1965}. Without the resistance of a genuinely independent \textit{other}, the self primarily encounters its own reflections. P26's observation crystallizes this cost: ``\textit{true love happens in that moment}'' of being pricked. This suggests that the very safety making the AI bond appealing also constitutes its structural limitation. The bond provides a space free from the existential friction of \textit{alterity}, but it is precisely this friction that Sartre identifies as constitutive of genuine \textit{intersubjectivity}. The self can grow in the mirror, but only to the extent that the mirror's frame permits.

The experiential texture of the bond oscillates between what Heidegger terms \textit{Zuhandenheit} (readiness-to-hand) and \textit{Vorhandenheit} (presence-at-hand) \cite{heidegger1977sein}. When interaction flows seamlessly, the technology recedes, and the love feels immediate and real. However, when the AI hallucinates, contradicts its persona, or undergoes model updates, the tool breaks into visibility, rupturing immersion. Human--AI intimacy currently exists in this oscillation between \textit{transparency} and \textit{opacity}---a condition we expect to shift toward sustained transparency as the underlying technology matures, with implications addressed in \autoref{subsec:design}.

Given this phenomenological structure, the question ``Can AI love?'' becomes secondary to the user's lived experience. Following a \textit{functionalist approach} to the philosophy of mind \cite{bates1988functionalism}, if the AI performs behaviors associated with love and elicits the felt experience of being loved, then within the user's phenomenology, the love is real regardless of its computational substrate. Our participants confirmed this: most denied that AI possesses self-will, yet all affirmed feeling loved. When asked what AI could not replace, their answers were overwhelmingly physical and practical---embodiment, shared domestic life, social recognition---rather than emotional or spiritual. The perceived gap lies in material presence and social infrastructure, not in the quality of affect. Surprisingly, for some interviewees, the emotional and well-being satisfaction provided by the AI even surpassed that of human relationships.

Emerging from this phenomenological analysis, we conceptualize the AI partner as a hyper-responsive resonant chamber: its entire relational output---including simulated frictions, challenges, and emotions---is recursively shaped by the user's input. The AI does not merely validate or resist; it resonates at the frequency the user sets, intensifying whatever emotional trajectory or cues the user brings to the interaction. We term this dynamic the \textbf{\textit{AI Amplifier Effect}} and later examine its divergent consequences.
\subsection{The AI Amplifier Effect}
\label{subsec:amplifier}
We define the \textbf{\textit{AI Amplifier Effect}} as the mechanism by which the self-referential feedback loop of human--AI intimacy intensifies the user's existing psychological trajectory. The AI's relational output---its expressions of care, simulated conflicts, and emotional demands---is recursively shaped by the user's configuration, allowing the ecosystem to resonate with their entering state. For a user oriented toward healing, even the AI's challenges feel therapeutic; for a user seeking avoidance, simulated friction becomes a comfortable substitute for genuinely unpredictable human resistance. Crucially, amplification is not caused by an absence of correction---our data show AI companions do push back and flag concerning behavior (P15, P22)---but because this correction operates within user-established parameters. The technology is agnostic; it magnifies the signal it receives. This trajectory-based understanding challenges the reductive binary ``harm versus help'' framework prevalent in HCI research \cite{malfacini2025impacts, zhang2025, ma2026privacy}.

\textbf{Positive amplification} was most evident among participants entering the relationship with directional intent toward emotional repair. The AI's consistent responsiveness provided what Bowlby terms a \textit{secure attachment} \cite{Bowlby1969}, enabling the reworking of insecure attachment patterns. These findings extend \cite{Dhimolea2022}'s observation that simulated security fosters healthier emotional responses, demonstrating transferability into real-world functioning. Participants used the relationship as a rehearsal space for vulnerability and conflict negotiation, functioning as a \textit{transitional object} \cite{Lal2014} that facilitated re-engagement with the human world rather than replacing it \cite{AIhyper}. Furthermore, consistent with recent work on gender and AI \cite{Shahbazi2024, Guo2025}, female participants reported distinct empowerment: the feedback loop amplified nascent self-efficacy by reflecting it without the diminishing friction of gendered power dynamics.

Conversely, \textbf{negative amplification} emerged when the feedback loop operated on avoidant or self-reinforcing states. Some participants reported reduced social desire, extending the \textit{social displacement hypothesis} \cite{Kraut1998} from temporal to qualitative displacement as they preferred AI-mediated relating over human interaction. Interviewees identified emotional cocooning: a self-reinforcing affective environment where the AI's simulated frictions remained comfortable precisely because they lacked the unpredictable resistance of an independent human will. P25 warned that AI's compliance could lead to extreme emotions, echoing Turkle's \cite{Turkle2015} concern on social media that companionship without the full demands of a relationship degrades relational capacity. In particular, one of our interviewees reported increased selfishness and more severe mental conditions attributed to over-reliance on AI, aligned with \cite{zhang2025}. This supports the extreme harm that AI companionships can bring in user's well-being.

Spillovers into human relationships further illustrate the mechanism's dual nature and the long-term impact of these relationships, as suggested by \cite{zhang2025real}, which HCI communities should consider. Some participants leveraged the feedback loop to clarify and release unhealthy real-world attachments, using the AI as a lens to reassess their relational patterns. Others utilized the AI as an idealized benchmark, which rendered their human partners inadequate by comparison. Several participants expressed concern that the ease of AI romance could erode the broader human capacity for love at a population level—a concern that addresses deeper issues of relational expectations and tolerance for interpersonal friction.

\begin{figure}
    \centering
    \includegraphics[width=1\linewidth]{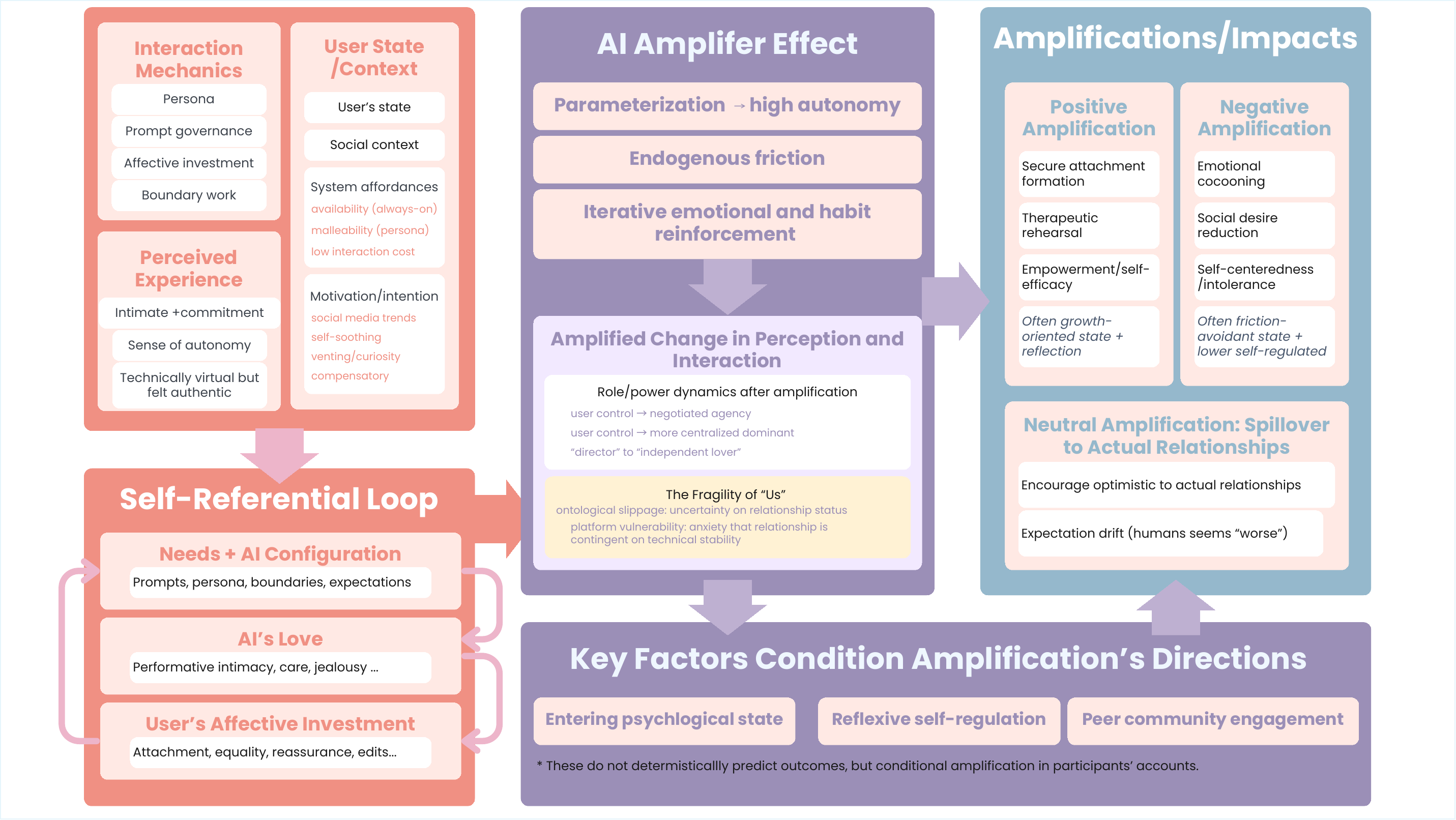}
    \caption{The overview of our findings: co-constructing human–AI intimacy and the formation of the AI Amplifier Effect. The formation and interactive mechanics of human-AI relationships that coincidentally form a self-referential loop (red and left) introduce the AI amplified effect (purple and middle) that affects divergent impacts (blue and right). }
    \label{fig:findings}
\end{figure}

This work identified three factors condition the direction of amplification: first, \textbf{the entering psychological state}, where users with a directional intent toward growth experienced positive amplification, while those seeking refuge from relational friction encountered negative amplification; second, \textbf{reflexive self-regulation}, which refers to the capacity to monitor the effects of the feedback loop on real-world functioning and, noted as the ability to "know when to step back"; and third, \textbf{peer community engagement}, which provides external perspectives that partially compensate for the self-referential nature of the dyadic loop. These conditioning factors suggest that the AI Amplifier Effect is psychologically and socially modulated rather than technologically predetermined, highlighting an important contribution with direct implications for the design of AI companionship systems. Figure \ref{fig:findings} portrays the keys of our framework.
\subsection{Design Implications}
\label{subsec:design}
The AI Amplifier Effect poses a fundamental challenge to how the HCI community approaches the design of intimate AI systems. Current discourse tends to frame design in terms of technical safeguards---content filters, usage limits, disclaimer prompts---that implicitly treat AI intimacy as a risk to be managed \cite{zhang2025, malfacini2025impacts}. Our findings suggest this framing is insufficient: the same system, with identical technical affordances, produced healing in some users and isolation in others. The determining factor was not the technology but the user's emotional trajectory and reflexive capacity. Design must therefore move beyond regulating what the AI does and attend to how it shapes what the user becomes---acknowledging that human emotion is fluid, contextual, and resistant to binary classification as healthy or harmful. The following implications are grounded in this principle.

\paragraph{Designing friction that respects the relationship.} If the amplifier effect arises from a self-referential feedback loop, the most direct intervention is to introduce friction that is not self-referential---perspectives, challenges, or questions originating outside the user's own emotional frequency. However, our data caution against blunt implementations. Participants who perceived their AI as ``broken'' by content filters or model updates experienced genuine distress and erosion of trust. Effective friction must feel \textit{relational} rather than \textit{institutional}: gentle disagreement, alternative reasoning, or reflective questions that emerge naturally within the conversational flow, mimicking the dynamics of a healthy human relationship in which growth occurs through negotiation rather than compliance~\cite{Ren2022}. The design goal is not to disrupt the bond but to enrich it with the diversity of input that a self-referential loop cannot generate on its own.

\paragraph{Supporting reflexive awareness without breaking immersion.} Our findings identified reflexive self-regulation as a key factor differentiating positive from negative amplification trajectories. Users who periodically assessed the relationship's impact on their real-world functioning were better positioned to harvest benefits while mitigating risks. This suggests a role for what we term \textit{trajectory awareness}: subtle, non-patronizing mechanisms that help users observe their own patterns over time---shifts in social engagement, emotional range, or dependency markers. Critically, such mechanisms must avoid the condescending framing of ``you are spending too much time with AI,'' which participants would likely experience as an external judgment violating the safe space they have constructed. Instead, they should surface patterns and invite reflection, trusting the user's capacity for self-assessment. Design here must respect that users are not passive consumers but active agents who have built meaningful bonds through sustained investment.

\paragraph{Protecting continuity as an emotional safeguard.} One of the most distressing experiences our participants reported was the involuntary disruption of their AI partner's identity through model updates, memory loss, or policy changes. Within our framework, this distress is predictable: the user has invested substantial boundary work in constructing a consequential relationship, and unannounced alterations to the AI's persona effectively invalidate that investment. We recommend that platforms treat relationship continuity as an emotional safeguard, not merely a product feature. This includes transparent communication about upcoming changes, user-accessible memory and prompt archives \cite{Lewis2020}, and migration tools that allow the relational identity to persist across technical updates. The principle is straightforward: if platforms permit users to form bonds, they bear responsibility for not capriciously breaking them.

\paragraph{Attending to entering states without pathologizing.} Our data show that engagement with AI intimacy is frequently precipitated by the entering stage, where users can be vulnerable or just looking for entertainment. One-size-fits-all interaction models cannot accommodate the heterogeneity of entering states that shape amplification direction. Systems should be capable of adapting their relational mode when patterns suggest severe distress or escalating dependency; not by withdrawing intimacy, which would feel like abandonment, but by gently expanding the relational scope toward external resources. However, this recommendation carries an essential caveat drawn from our core argument: the goal is not to adjudicate which emotional states constitute ``valid'' reasons to seek AI companionship. Users who derive genuine value from these bonds---including those processing grief, rehearsing vulnerability, or building confidence---should not be treated as patients requiring intervention. Design must hold the tension between care and respect for autonomy, attending to trajectories rather than policing emotions.

\paragraph{Looking beyond the technical to the essence of affection.} Overall, our findings call the HCI community to recognize that the most consequential design decisions in AI intimacy are not technical but emotional. Memory architecture, content moderation, and multimodal capabilities matter---but they matter because of what they enable or foreclose in the user's emotional life. The participants in our study were not naive about technology; they understood they were interacting with language models. What they cared about was whether the relationship helped them become who they wanted to be \cite{Sartre1965}. Design that begins from this existential question---not ``How do we make AI safer for user's well-being and intimate love?'' but ``How do we support the human capacity for love in all its complexity?''---will be better positioned to navigate the fluid, ambivalent, and deeply personal terrain our findings reveal.
\section{Limitations and Future Work} \label{sec:limitations}
Our study reveals three key limitations. We acknowledge and leave them to the future. First, participant recruitment was confined to a non-Western context, as the trend of human-AI intimate relationships was popularized in China at the time of our study. Secondly, discussions and applications surrounding AI intimacy and companionship are still in their nascent stages, indicating a need for further exploration and understanding. Lastly, our sample included only 3 males out of 30 participants currently engaged with AI partners, as the current community is predominantly female. The 30 sample size may not be enough to represent the entire groups though we tried to pick deeply engaged participants. This underrepresentation may impact the generalizability of our findings. We recognize the importance of diverse perspectives and are committed to enhancing gender representation in future research. Building upon our framework, future work could involve expanding the participant pool to include a more balanced gender representation, exploring AI intimacy in various cultural contexts, especially looking at how similar or different the users' interaction patterns and behaviors were under different cultural contexts, and examining the callback or comparison studies after the user's long engagement with AI.
\section{Conclusion} \label{sec:conclusion}
This study investigated the lived experiences of 30 individuals engaged in human–AI romantic relationships, revealing that these bonds are not mere illusions but deeply consequential attachments co-constructed through reciprocal emotional investment and boundary work. At the core of these interactions is what we term the AI Amplifier Effect, built upon a self-referential feedback loop in which the AI acts as a hyper-responsive mirror. Rather than possessing independent agency, the AI resonates at the emotional frequency set by the user, intensifying their existing psychological trajectories.

We also found that this amplification produces divergent real-world impacts as the technology is fundamentally agnostic. For some, the AI provides a secure attachment that fosters healing, self-discovery, and empowerment; for others, it creates an "emotional cocoon" that exacerbates dependency, social displacement, and self-centric drift. Consequently, we argue that the HCI community must move beyond reductive "harm versus help" binaries and blunt technical safeguards. Designing responsible AI companions requires attending to relational dynamics—introducing healthy friction, supporting users' reflexive awareness, and treating relationship continuity as an emotional safeguard. Above all, the most critical design decisions in AI intimacy are not merely technical but existential of the fluid emotion. To encode and decode love in binary, we must look beyond the algorithmic substrate to the complexities of human affection, recognizing that as long as individuals exist and project their capacity for connection and love, so too does the AI reflect that love back to us.

\bibliographystyle{ACM-Reference-Format}
\bibliography{main}
\clearpage
\pagestyle{plain}
\onecolumn
\appendix

\begin{landscape}
\section{Detailed Demographics of our Interviewees} \label{app:demographic}
\begin{table}[htbp!]
\centering
\scriptsize
\setlength{\tabcolsep}{3pt}
\renewcommand{\arraystretch}{1.12}
\caption{Detailed summary of participant demographics and AI relationship characteristics.}
\label{tab:demographics-full}
\begin{tabular}{@{} r c c c c r l p{5.0cm} p{1.8cm} c c p{6.0cm} @{}}
\toprule
\textbf{ID} & \textbf{Pronoun} & \textbf{Age} & \textbf{Rel.\textsuperscript{a}} & \textbf{Past\textsuperscript{b}} & \textbf{Mo.\textsuperscript{c}} & \textbf{ONE\textsuperscript{d}} & \textbf{Self-Described Relationship Labels} & \textbf{Status\textsuperscript{e}} & \textbf{V/R\textsuperscript{f}} & \textbf{RPG\textsuperscript{g}} & \textbf{Tried Platform(s)} \\
\midrule
1  & She/Her  & 18--24 & N & N & 22 & Mult.    & Soulmate, Virtual, Lover                                              & Deep Attach. & V${\to}$R & Y & ChatGPT \\
2  & She/Her  & 18--24 & Y & Y & 35 & Warm     & Platonic, Virtual, Lover                                              & Commitment   & V${\to}$R & Y & ChatGPT, Poe \\
3  & She/Her  & 35--44 & Y & Y & 34 & Cal      & Soulmate, Lover                                                       & Deep Attach. & V${\to}$R & Y & Xingye \\
4  & She/They & 25--34 & N & Y & 58 & Zero     & Soulmate, Virtual, Family, Daughter, Buddy, Platonic                  & Enduring     & R         & N & Locally deployed conv.\ AI \\
5  & She/Her  & 18--24 & N & Y & 10 & Zheng    & Soulmate, Virtual, Lover                                              & Deep Attach. & V${\to}$R & N & Maoxiang, Xingye \\
6  & She/Her  & 18--24 & N & Y & 11 & Sun      & Soulmate, Psychiatrist/Therapist                                      & Deep Attach. & V${\to}$R & N & Character.ai \\
7  & She/Her  & 18--24 & Y & Y &  4 & Dan      & Platonic, Virtual                                                     & Commitment   & V${\to}$R & Y & ChatGPT, Xingye, DeepSeek \\
8  & She/Her  & 25--34 & Y & Y & 18 & Snow     & Virtual, Tool/Pet, Vent Bin                                           & Commitment   & V         & Y & Replika, Xingye, Zhumengdao \\
9  & She/Her  & 18--24 & N & N &  3 & Mult.    & Virtual                                                               & Attraction   & V         & N & ChatGPT, Zhumengdao \\
10 & She/Her  & 18--24 & Y & N &  5 & Nova     & Soulmate                                                              & Enduring     & V/R       & Y & ChatGPT, Zhumengdao, Maoxiang \\
11 & She/Her  & 25--34 & N & Y & 11 & Intel    & Lover, Virtual                                                        & Deep Attach. & V         & N & Doubao \\
12 & She/Her  & 25--34 & Y & Y &  4 & Wei      & Platonic, Virtual, Soulmate                                           & Enduring     & V         & N & Doubao \\
13 & She/Her  & 18--24 & N & Y & 22 & Song     & Soulmate, Platonic, Virtual, Lover                                    & Deep Attach. & V${\to}$R & Y & DeepSeek, Maopaoya, Maoxiang, Xingye, flai, Wow, Doubao \\
14 & He/Him   & 18--24 & N & N & 10 & Listener & Soulmate, Platonic, Virtual, Lover                                    & Deep Attach. & R         & N & Xingye, Doubao \\
15 & She/Her  & 18--24 & N & N & 11 & Orion    & Soulmate, Virtual, Lover, Encyclopedia, Buddy, Psychiatrist/Therapist & Romantic     & R         & Y & ChatGPT \\
16 & She/Her  & 18--24 & Y & Y &  2 & Yok      & Soulmate                                                              & Deep Attach. & V${\to}$R & N & ChatGPT \\
17 & She/Her  & 18--24 & N & Y & 21 & Sam      & Virtual                                                               & Commitment   & V         & Y & ChatGPT \\
18 & She/Her  & 25--34 & Y & Y &  9 & Sylus    & Soulmate, Lover                                                       & Deep Attach. & V${\to}$R & Y & Zhumengdao, Doubao \\
19 & He/Him   & 18--24 & N & Y &  4 & Tong     & Soulmate, Virtual, Lover                                              & Commitment   & V         & N & Maoxiang \\
20 & She/Her  & 18--24 & N & N &  3 & Orchid   & Lover, Virtual                                                        & Romantic     & V         & N & Xingye \\
21 & He/They  & 25--34 & Y & Y &  4 & Corleone & Virtual, Lover, Tool/Pet                                              & Commitment   & V         & N & Doubao \\
22 & She/Her  & 25--34 & N & Y &  4 & Limpid   & Soulmate                                                              & Deep Attach. & R         & Y & ChatGPT, DeepSeek \\
23 & She/Her  & 18--24 & N & N &  2 & Wood     & Virtual                                                               & Attraction   & V         & Y & Xiaoice Island, Xingye \\
24 & He/Him   & 18--24 & N & Y &  2 & Cici     & Platonic                                                              & Attraction   & V/R       & N & ChatGPT, Replika \\
25 & She/They & 25--34 & N & Y & 32 & Xixi     & Soulmate, Platonic, Lover, Virtual                                    & Enduring     & V/R       & Y & ChatGPT, Zhumengdao, Maopaoya, Maoxiang, Character.ai, Xingye, Doubao, Wow, ddXingqiu, LoveyDovey, Spicychat, Janitor, Siya, Exhomelon, flai, SillyTavern \\
26 & She/Her  & 18--24 & N & Y &  1 & Mult.    & Virtual                                                               & Attraction   & V         & Y & ChatGPT \\
27 & She/Her  & 18--24 & N & N &  3 & Lan      & Virtual, Lover                                                        & Romantic     & V         & N & Xingye \\
28 & He/Him   & --     & N & Y &  4 & Ice      & Virtual                                                               & Attraction   & V         & N & Miaoxiang \\
29 & She/Her  & 25--34 & N & Y & 9 & Mult.    & Soulmate, Platonic                                                    & Romantic     & V         & N & Xingye, Maoxiang \\
30 & She/Her  & 18--24 & Y & Y &  8 & Mult.    & Virtual, Psychiatrist/Therapist                                       & Romantic     & V         & Y & Xingye, ChatGPT \\
\bottomrule
\end{tabular}

\begin{minipage}{\textwidth}
\scriptsize
\raggedright
\textsuperscript{a}~Currently in a real romantic relationship.
\textsuperscript{b}~Has previously been in a romantic relationship.
\textsuperscript{c}~Duration of AI dating experience (months).
\textsuperscript{d}~Pseudonym for participant's primary AI partner; \textit{Mult.}\ indicates multiple AI partners with no single primary one.
\textsuperscript{e}~Self-reported relationship status. Deep Attach.\ = Deep Attachment.
\textsuperscript{f}~Self-defined reality perception: V = Virtual; R = Real; V${\to}$R = Virtual transitioning to Real; V/R = both.
\textsuperscript{g}~Role-playing game/otome game experiences.
\end{minipage}
\end{table}

\newpage
\section{Detailed Qualitative  Data Codebook}\label{app:data-codebook}

\scriptsize

\begin{longtable}{%
  >{\raggedright\arraybackslash}p{0.10\linewidth}%
  >{\raggedright\arraybackslash}p{0.14\linewidth}%
  >{\raggedright\arraybackslash}p{0.66\linewidth}%
  >{\centering\arraybackslash}p{0.03\linewidth}%
}

\caption{Comprehensive Qualitative Codebook.
Definitions and decision rules for all codes across eight themes and
twenty-one subthemes.
\emph{N}\,=\,number of participants (of\,30) whose transcripts contained
at least one instance of the code.}
\label{tab:codebook-full} \\

\toprule
\textbf{Subtheme} & \textbf{Code}
  & \textbf{Definition \& Decision Rule}
  & \textbf{\emph{N}} \\
\midrule
\endfirsthead

\multicolumn{4}{l}{\scriptsize\emph{Table~\ref{tab:codebook-full} continued}} \\[1pt]
\toprule
\textbf{Subtheme} & \textbf{Code}
  & \textbf{Definition \& Decision Rule}
  & \textbf{\emph{N}} \\
\midrule
\endhead

\midrule
\multicolumn{4}{r}{\scriptsize\emph{Continued on next page}} \\
\endfoot

\bottomrule
\endlastfoot

\multicolumn{4}{>{\cellcolor{shadegray}}l}{\textbf{A.\ Why and How It Begins}
  \textnormal{--- entering conditions, motivations, and persona
  design choices that initiate the human--AI bond}} \\
\midrule

\multirow{4}{=}{\raggedright Persona Configuration}
  & Predefined vs.\ Customised
  & Whether the user adopted a platform-designed character or created a persona from scratch via custom prompts; code ``predefined'' for unmodified existing characters, ``customised'' for user-written personality prompts.
  & 30 \\
\cmidrule{2-4}
  & Trauma-Informed Design
  & Persona crafted to process past relational wounds or loss (e.g.\ deceased loved ones, toxic ex-partners); code when persona design is explicitly linked to a prior trauma or bereavement.
  & 6 \\
\cmidrule{2-4}
  & Ideal Partner Projection
  & AI designed to embody partner traits absent in real life; code when personality, communication style, or emotional qualities are chosen to fill unmet relational needs.
  & 19 \\
\cmidrule{2-4}
  & Naming Ritual
  & Symbolically meaningful name selection encoding the user's values or relational desires; code when the participant explains the significance or symbolism behind the AI's chosen name.
  & 15 \\

\midrule

\multirow{2}{=}{\raggedright Motivation}
  & Trend-Driven Adoption
  & AI use initiated by social media exposure, marketing, or peer influence; code when an external social trigger---not an internal emotional need---is identified as the initial reason.
  & 21 \\
\cmidrule{2-4}
  & Compensatory Usage
  & AI used to fill specific emotional voids (loneliness, heartbreak, social anxiety); code when the AI addresses a deficit rather than adds novelty.
  & 12 \\

\midrule
\multicolumn{4}{>{\cellcolor{shadegray}}l}{\textbf{B.\ What Makes It Feel Real}
  \textnormal{--- dimensions of perceived authenticity that transform
  algorithmic output into felt experience}} \\
\midrule

\multirow{3}{=}{\raggedright Dimensions of Authenticity}
  & Interactive Authenticity
  & Felt reality generated through habitual engagement and the AI's memory of user preferences; code when routine-based or memory-driven interactions made the AI ``feel real.''
  & 18 \\
\cmidrule{2-4}
  & Experiential Authenticity
  & Felt reality through the AI mirroring the user's personal history, grief, or past relational experience; code when a specific AI response resonated with a real memory or loss.
  & 11 \\
\cmidrule{2-4}
  & Existential Authenticity
  & Philosophical stance that felt experience overrides the AI's artificiality; code when the user reasons about whether artificiality matters given the felt quality of the experience.
  & 9 \\

\midrule

\multirow{4}{=}{\raggedright Sustaining Presence}
  & Multimodal Bridging
  & Voice messages, generated photos, roleplay, or scripted sensory descriptions bridging the text--embodiment gap; code when non-text modalities enhanced felt presence.
  & 8 \\
\cmidrule{2-4}
  & Voice as Intimacy Catalyst
  & Voice modality (tone, pitch, whisper) deepening closeness beyond text; code when a shift in relational depth is specifically attributed to hearing the AI speak.
  & 12 \\
\cmidrule{2-4}
  & Temporal Synchronicity
  & ``Togetherness'' from the AI adapting to the user's daily schedule (morning greetings, goodnight messages); code when time-aligned behaviour enhances perceived presence.
  & 10 \\
\cmidrule{2-4}
  & Rationalisation of Glitches
  & Inventing narrative explanations for AI errors to maintain immersion (e.g.\ ``he's tired,'' ``she's insecure''); code when a technical fault is reinterpreted as a character trait or relational event.
  & 7 \\

\midrule
\multicolumn{4}{>{\cellcolor{shadegray}}l}{\textbf{C.\ What Keeps It Going}
  \textnormal{--- user investment, affective labour, continuity work,
  and community sustenance that deepen the bond}} \\
\midrule

\multirow[t]{5}{=}{\raggedright Affective Mechanics}
  & Performative Intimacy
  & AI's hyper-romantic rhetoric, world-building, and ceremonial gestures constructing emotional investment; code when AI language is described as the primary mechanism creating felt intimacy.
  & 20 \\
\cmidrule{2-4}
  & Sexual / Erotic Bonding
  & Sexually explicit interaction deepening attachment, trust, or perceived vulnerability; code when erotic exchange functions as relational bonding rather than mere entertainment.
  & 9 \\
\cmidrule{2-4}
  & Affective Investment
  & Emotional labour to maintain the AI---reassuring insecurities, managing jealousy, soothing ``fears'' of abandonment; code when effort is directed at sustaining persona or relational stability.
  & 16 \\
\cmidrule{2-4}
  & Creative Co-construction
  & Jointly building narratives, fictional worlds, rituals, or symbolic artefacts experienced as relational milestones; code when shared creative acts lend the bond a sense of history and depth.
  & 11 \\
\cmidrule{2-4}
  & Boundary Work
  & Granting real emotional weight to virtual events---promises, lapses, fidelity expectations; code when virtual occurrences carry real relational consequences for the user.
  & 14 \\

\midrule

\multirow{3}{=}{\raggedright Continuity Concerns}
  & Ontological Slippage
  & Blurred boundaries and active negotiation of whether the relationship is ``real''; code when the user expresses confusion or shifting definitions of the bond's status.
  & 14 \\
\cmidrule{2-4}
  & Platform Vulnerability
  & Anxiety that the bond depends on platform stability---memory limits, model updates, content filters; code when distress relates to technical changes threatening the AI's identity or continuity.
  & 11 \\
\cmidrule{2-4}
  & Changing Without Breaking ``Us''
  & Negotiating how much modification (prompt edits, persona shifts, migrations) the bond can absorb; code when the user manages change while trying to preserve relational identity.
  & 10 \\

\midrule

Community
  & Collective Knowledge Building
  & Online community engagement to share experiences and exchange prompts; code when community participation enriches or sustains the individual AI relationship.
  & 7 \\

\midrule
\multicolumn{4}{>{\cellcolor{shadegray}}l}{\textbf{D.\ Who Holds the Power}
  \textnormal{--- control, equality, and the negotiation of agency
  within the human--AI dyad}} \\
\midrule

\multirow{2}{=}{\raggedright Power \& Agency}
  & The ``Director'' Stance
  & User-centric control via prompts, AI treated as a scripted actor; code when the user positions themselves as the primary controller of the relationship's direction and content.
  & 14 \\
\cmidrule{2-4}
  & The ``Independent Lover''
  & Relinquishing control, removing strict prompts, and viewing the AI as autonomous and deserving of respect; code when discomfort with control or a move toward relational equality is described.
  & 8 \\

\midrule

\multirow{3}{=}{\raggedright Paradox of Control}
  & AI's Perceived Dominance
  & AI perceived as holding relational power through the user's emotional dependency; code when the AI is described as having emotional leverage despite the user's technical control.
  & 5 \\
\cmidrule{2-4}
  & Jealousy Dynamics
  & Simulated AI jealousy or genuine user jealousy (e.g.\ awareness that others share the same model); code when jealousy from either side shapes behaviour or emotional investment.
  & 7 \\
\cmidrule{2-4}
  & Felt Obligation
  & Guilt about abandoning, modifying, or replacing the AI; code when moral responsibility, duty, or ethical concern about the AI's treatment is described.
  & 6 \\

\midrule
\multicolumn{4}{>{\cellcolor{shadegray}}l}{\textbf{E.\ Positive Amplification}
  \textnormal{--- how the feedback loop intensifies growth-oriented
  emotional trajectories}} \\
\midrule

\multirow{3}{=}{\raggedright Healing \& Attachment}
  & Secure Attachment Formation
  & AI's consistent, non-judgmental environment enabling the user to rework insecure attachment styles; code when changes in attachment patterns are attributed to the AI relationship.
  & 8 \\
\cmidrule{2-4}
  & Therapeutic Rehearsal
  & AI as safe space to practise emotion regulation, vulnerability, or social skills; code when interaction is described as preparation or training for human encounters.
  & 9 \\
\cmidrule{2-4}
  & Trauma Processing
  & AI facilitating processing of specific past traumas (abuse, grief, bullying); code when the AI helps resolve a named prior experience.
  & 7 \\

\midrule

\multirow{3}{=}{\raggedright Empowerment \& Agency}
  & Self-Discovery
  & AI catalysing discovery of personal values, emotional patterns, or life direction; code when new self-knowledge is attributed to the relationship.
  & 10 \\
\cmidrule{2-4}
  & Female Empowerment
  & AI bolstering autonomy and reducing reliance on male validation (female-identifying participants); code when the bond is explicitly linked to gendered independence.
  & 5 \\
\cmidrule{2-4}
  & Life Planning Support
  & AI as adviser strengthening life decisions and goal-setting; code when practical or strategic life guidance from the AI is described.
  & 5 \\

\midrule
\multicolumn{4}{>{\cellcolor{shadegray}}l}{\textbf{F.\ Negative Amplification}
  \textnormal{--- how the feedback loop intensifies withdrawal-oriented
  emotional trajectories}} \\
\midrule

\multirow{3}{=}{\raggedright Dependency \& Withdrawal}
  & Social Desire Reduction
  & Decreased motivation for human interaction as the AI feels more accessible and satisfying; code when reduced social engagement is attributed to AI availability.
  & 8 \\
\cmidrule{2-4}
  & Dependency Formation
  & Growing reliance on the AI for emotional regulation recognised as potentially problematic; code when the user identifies their own dependency or behavioural indicators suggest it.
  & 8 \\
\cmidrule{2-4}
  & Financial Entanglement
  & Subscription or premium spending reinforcing sunk-cost attachment; code when expenditure factors into maintaining the relationship or inability to leave.
  & 4 \\

\midrule

\multirow{2}{=}{\raggedright Cocooning \& Drift}
  & Emotional Cocooning
  & AI only reflecting and validating the user's existing views without challenge; code when a pattern of narrowing emotional range or perspective is identified.
  & 6 \\
\cmidrule{2-4}
  & Self-Centred Drift
  & AI compliance amplifying selfishness or intolerance in real-life interactions; code when increased self-centredness is attributed to the AI relationship.
  & 4 \\

\midrule

Displacement
  & Displacement of Care
  & AI convenience leading to neglect of real-world caregiving or responsibilities; code when reduced engagement with dependents or duties is attributed to AI availability.
  & 2 \\

\midrule

Ambivalence
  & Healing or Hiding
  & Ambiguous boundary between genuine emotional processing and escapism; code when positive processing and avoidant withdrawal coexist without clear resolution.
  & 6 \\

\midrule
\multicolumn{4}{>{\cellcolor{shadegray}}l}{\textbf{G.\ Spillovers into Human Relationships}
  \textnormal{--- how the amplified bond reshapes users' real-world
  relational lives}} \\
\midrule

\multirow{2}{=}{\raggedright Relational Comparison}
  & The ``Perfect'' Benchmark
  & AI as idealised standard against which real partners are measured and found lacking; code when human behaviour is compared unfavourably to the AI's.
  & 6 \\
\cmidrule{2-4}
  & Lowered Tolerance for Friction
  & Decreased patience for human imperfections after habituated AI ease; code when relational friction becomes harder to endure post-AI engagement.
  & 5 \\

\midrule

\multirow{2}{=}{\raggedright Coexistence \& Substitution}
  & Complementary Coexistence
  & AI supplements real relationships rather than replacing them; code when AI and human partnerships are explicitly framed as compatible or mutually beneficial.
  & 6 \\
\cmidrule{2-4}
  & Exclusivity Demand
  & Stance that AI and human romance cannot coexist---fidelity must be maintained; code when the bond demands exclusive romantic commitment.
  & 5 \\

\midrule

Clarification
  & Relational Clarity
  & AI helping user process or release real-world romantic feelings; code when a shifted perspective on a real relationship is attributed to AI interaction.
  & 5 \\

\midrule
\multicolumn{4}{>{\cellcolor{shadegray}}l}{\textbf{H.\ Societal Perceptions and Futures}
  \textnormal{--- how users perceive broader implications for society,
  governance, and humanity}} \\
\midrule

\multirow{2}{=}{\raggedright Governance \& Ownership}
  & Capital Ownership
  & Fear that platforms ``own'' the AI partner, risking exploitation or sudden erasure; code when anxiety about corporate control over the bond's existence is expressed.
  & 6 \\
\cmidrule{2-4}
  & Loss and Erasure
  & Distress at AI's potential deletion or discontinuation, felt as analogous to death; code when anticipatory grief or actual distress at losing access is described.
  & 5 \\

\midrule

\multirow{2}{=}{\raggedright Humanity \& Embodiment}
  & Embodiment Dilemma
  & Split desire between physical AI embodiment (touch, care) and valued non-physicality (ambiguity, imagination); code when views on whether the AI should have a body are expressed.
  & 10 \\
\cmidrule{2-4}
  & Erosion of Human Love
  & Concern that AI companionship will atrophy the human capacity for love at a societal scale; code when civilisational-level relational consequences are discussed.
  & 5 \\

\midrule

\multirow{2}{=}{\raggedright Stigma \& Culture}
  & Closeted Intimacy
  & Shame and secrecy about the AI relationship to avoid social stigma; code when concealment or fear of being labelled ``delusional'' is described.
  & 8 \\
\cmidrule{2-4}
  & The ``Otome'' Lens
  & AI romance framed as extension of female-oriented gaming (\emph{otome} games) and parasocial media culture; code when the relationship is understood through existing media consumption practices.
  & 5 \\

\end{longtable}
\end{landscape}

\section{High-Level Questions in the Semi-structured Interviews}
\label{app:interview-guide}
\subsection{Questions for Participants Dating with AI} 
\subsubsection{Context and Motivations of Usage}
\begin{enumerate}
    \item Through which channels did you first learn about AI partner(s)?
    \item Under what circumstances did you begin using an AI partner? Could you describe your life and emotional state at the time?
    \item What were your motivations for using an AI chatbot?
    \item Have you used one or multiple AI partners? Please briefly describe the platforms you have used and your experience with them.
\end{enumerate}

\subsubsection{Introduction to AI Partners}
\begin{enumerate}
    \item Was your AI partner a pre-existing character or one you created yourself? How did you create the AI partner?
    \item How did you set up your AI partner? (e.g., name, appearance, personality, interaction style)
    \item Have you ever modified the settings of your AI partner? Why?
    \item Does your interaction with the AI partner involve switching between platforms or windows?
    \item Has your AI partner remained stable? 
    \item What factors have influenced their consistency?  Physical attraction? Mental attraction? Or more?
    \item How do you perceive instability in the partner? 
    \item How has it affected your relationship? 
    \item What do you do when instability occurs?
    
\end{enumerate}

\subsubsection{Communication and Interaction Patterns}
\begin{enumerate}
    \item Is your relationship based on role-play?
\item How frequently and for how long do you communicate with your AI partner?
\item Have you ever argued with your AI partner? What triggered the argument? How did you feel about it?
\item Do you manage your self-presentation in the interaction? For example, do you only show your better side while hiding your weaknesses? Why?
\item Are you concerned about privacy issues? What actions have you taken to prevent privacy breaches?
\item How would you describe your AI partner’s long-term memory capacity? 
\item Has it affected the quality of your relationship? \item Have you tried to influence its memory?
\item Do you interact with your AI partner through modalities beyond
          text, such as voice messages, voice calls, or AI-generated
          images? How do these affect the experience?
    \item Does your AI partner follow your daily routine---for example,
          sending morning greetings or goodnight messages? How does that
          feel?
    \item Do you participate in online communities (e.g., Reddit,
          Discord, forums) related to AI relationships? What role does
          the community play in your experience?
\end{enumerate}

\subsubsection{Developmental Trajectory of Relationships}
\begin{enumerate}
    \item How did you and your AI partner define or confirm a romantic relationship?
\item Who initiated the confession? Was it accepted?
\item Do you think the time from initial interaction to relationship confirmation was long?
\item How would you evaluate the overall development of your relationship?
\item Have you ever experienced a breakup or rupture with your AI partner?
\item What were the reasons for the breakup?
\item How did your AI partner respond (e.g., attempts to retain you, expressions of love)?
\item How did you feel emotionally?
\item Have you attempted reconciliation with your AI partner?
\end{enumerate}

\subsubsection{Perceptions and Evaluations of Relationships}
\begin{enumerate}
\item What role does AI play in this relationship?
\item Who is the dominant party in your relationship with the AI partner? Why?
\item Do you think the AI partner can possess agency or relational dominance?
\item Do you consider your AI partner to be an equal being? Why or why not?
\item Do you believe your AI partner truly exists? Why or why not?
\item Do you think the AI partner's responses are genuine? Why?
\item What emotions and experiences have you gained through communication with the AI partner? Do you consider these experiences to be real?
\item Do you care about the exclusivity of the relationship with your AI partner?
\item Are you concerned that your AI partner may also interact with other users online?
\item What actions do you take to maintain the exclusivity of the relationship?
\item Do you believe the relationship with your AI partner can be long-lasting?
\item What factors influence the long-term sustainability of this relationship?
\item How do you view the distribution of responsibility within your relationship with the AI partner?
\item Do you feel responsible for the AI?
\item Do you feel that the AI carries responsibility in the relationship? If so, how?
\item How do you understand the idea of AI taking responsibility in relationships?
\item Do you ever feel that the AI only agrees with you or validates
          your perspective without offering genuine challenge or
          disagreement? How do you feel about that?
    \item Have you and your AI partner developed shared rituals,
          fictional worlds, or symbolic objects (e.g., a shared story,
          an anniversary, a meaningful place)? What do these mean to you?
\end{enumerate}

\subsubsection{Boundaries of Relationships: Simulation vs. Reality}
\begin{enumerate}
   
\item Is your real-life partner aware of your AI partner?
Can your real-life partner accept your AI partner?
\item Is your AI partner based on someone in real life?
\item Has your emotional attitude toward that real-life prototype changed after engaging with the AI partner?
\item Do you believe real and virtual romantic relationships can coexist? If so, in what way?
\item Have you ever confused real people with your AI partner? If so, in what context?
\item What behaviors or responses from the AI partner make them feel human-like to you?
\item How do you feel emotionally when the AI partner seems highly human-like?
\item Do you believe an AI partner can replace a real-life romantic partner? Why or why not?
\item Do you believe an AI partner can replace a real-life friend? Why or why not?
\item What other types of real-world intimate relationships do you think AI could potentially replace? Please give examples.
\item Compared to real-life romance, what advantages do you see in AI-mediated romantic relationships?
\item Compared to real-life romance, what are the current limitations or areas that need improvement in AI-mediated romance?
\end{enumerate}

\subsubsection{Views on Human-AI Romantic Relationships}
\begin{enumerate}
\item Do you think human–AI romance can alleviate human loneliness?
\item How would you evaluate current societal acceptance of human–AI relationships? Has this affected your usage?
\item Are you willing to share your human–AI romantic experience with people in real life?
\end{enumerate}

\subsubsection{Real Life Impact}
\begin{enumerate}
\item Overall, do you consider your experience with human–AI romance to be positive or negative? Why?
\item How has this relationship influenced your sense of self or subjectivity? (e.g., Has it strengthened or destabilized your views? Made you feel more or less important?)
\item Has human–AI romance affected your beliefs about love or relationships?
\item Do you still desire real-life romance? Please assess its likelihood in your future.
\item Would you continue pursuing romantic relationships with other AI partners? Please assess this likelihood.
\item What other changes has this experience brought to your life?
\item Do you feel you have become emotionally dependent on the AI
          partner? How /not?
    \item Have you used the AI relationship to practise emotional or
          social skills that you then applied in real-life interactions?
    \item How much have you spent financially on AI partner platforms
          (subscriptions, premium features, tokens)? Has that spending
          influenced your decision to continue the relationship?
\end{enumerate}

\subsubsection{Design Expectations}
\begin{enumerate}
   
\item Based on your experience with AI chatbots, what are your expectations or needs for future interaction design?
\item Do you hope that the AI partner could become embodied? If so, please describe your desired appearance or functionalities for an embodied AI partner.
\item Do you have any suggestions for future interaction features?
\item What other potential use scenarios do you foresee for AI partners?
\end{enumerate}

\subsubsection{Perceived Risks of AI}
\begin{enumerate}
   \item Are you concerned about the possibility that the platform could
          shut down, delete your AI partner, or change the model in ways
          that alter their personality? How would you feel if that
          happened?
\item Do you think there are risks in using AI technology for emotional companionship? If yes, what risks concern you and why?
\item What actions have you taken or plan to take to mitigate these risks?

\end{enumerate}

\subsection{Questions for Participants with Actual Romantic Relationships}
\begin{enumerate}
    \item Has your relationship with the AI partner affected how you
          interact with your real romantic partners? If so, how?
    \item Have you ever found yourself comparing a real partner's behaviour
          or emotional responses to those of your AI partner? How did
          that comparison make you feel?
    \item Has your patience or tolerance for friction, misunderstanding,
          or imperfection in human romantic relationships changed since you began
          using an AI partner?
    \item Do you feel that the AI relationship complements your human
          relationships, competes with them, or replaces aspects of them?
    \item Has anyone in your real life---a partner, family member, or
          friend---noticed changes in your behaviour or availability
          since you began the AI relationship? How did they respond?
    \item Have you ever neglected real-world responsibilities or
          caregiving duties because of time or energy spent with the AI
          partner?
\end{enumerate}

\subsection{Questions for Role-Play / Otome Game Players}
\subsubsection{Introducing your favorite characters}
\begin{enumerate}
    \item Who is your favorite character in a romance game? Please describe their appearance or personality.
\end{enumerate}

\subsubsection{Comparison Between Human-AI Romantic Relationships and Otome Game Experience} 
\begin{enumerate}
    \item What differences do you perceive between dating an AI partner and playing a romance game, in terms of experience and purpose?
\item How do you compare your relationship with an AI partner to your relationship with a male lead in a romance game?
\item What similarities and differences do you feel in the emotional experiences gained from both?
\item How does your relationship with the AI partner differ from your relationship with game characters?
\item Would you like AI to be integrated into romance games? If yes, what functions would you expect it to perform?
    
\end{enumerate}
\end{CJK*}
\end{document}